%% file: NewOldIdea.tex
\documentclass[a4,journal]{IEEEtran}
\usepackage{amsmath,amssymb,dsfont,stfloats,color,url,mathtools}
\usepackage[pdftex]{graphicx}
\usepackage{wasysym,esint}
\usepackage{cite}
\usepackage{caption}
\usepackage{subcaption}
\usepackage{gensymb} 
\usepackage[boxruled,linesnumbered]{algorithm2e}
\usepackage{algorithm2e}
\usepackage{lipsum}

\DeclareGraphicsExtensions{.eps,.pdf,.png,.jpg,.gif,.jpeg,.pstex}

\newtheorem{rem}{Remark}

\input{macros}

\title{A New Old Idea: Beam-Steering Reflectarrays for Efficient Sub-THz Multiuser MIMO}

\author{\IEEEauthorblockN{Krishan Kumar Tiwari\IEEEauthorrefmark{1}, \emph{Senior Member, IEEE}, and Giuseppe Caire\IEEEauthorrefmark{1}, \emph{Fellow, IEEE}} \thanks{\IEEEauthorrefmark{1}Technical University of Berlin, Germany. Email ids: lastname@tu-berlin.de}}

\begin{document}

\bstctlcite{BSTcontrol} 

\maketitle

\begin{abstract}
This paper presents a novel, power- and hardware-efficient, multiuser, multibeam RIS (Reflective Intelligent Surface) architecture for multiuser MIMO, 
especially suited to operate in very high frequency bands (e.g., high mmWave and sub-THz), where channels are typically sparse in the beamspace and line-of-sight (LOS)  is the dominant component. The key module is formed by an active multiantenna feeder (AMAF) with a small number of active antennas, placed in the near field of a RIS with a much larger number of passive controllable reflecting elements. We propose a pragmatic approach to obtain 
a steerable beam with high gain and very low sidelobes. Then $K$ independently controlled beams can be achieved by closely stacking $K$ such 
AMAF-RIS modules. Our analysis includes the mutual interference between the modules and the fact that, due to the delay difference of propagation through
the AMAF-RIS structure, the resulting channel matrix is frequency selective even in the presence of pure LOS propagation. 
We consider a 3D geometry and show that ``beam focusing'' is in fact possible (and much more effective in terms of coverage) 
also in the far-field, by creating spotbeams with limited footprint both in angle and in range. 
Our results show that: 1)  simple RF beamforming without computationally expensive baseband digital multiuser precoding
is sufficient to practically eliminate multiuser interference when the users are chosen with sufficient angular/range separation, thanks to the extremely low sidelobes of the proposed module; 2) the impact of beam pointing errors with standard deviation as large as 2.5 deg 
and RIS quantized phase-shifters with quantization bits $\geq 3$ is essentially negligible; 3) The proposed architecture is more power efficient and much simpler from a hardware implementation viewpoint than standard RF beamforming {\em active arrays} with the same beamforming performance. 
As a side result, we show also that the array gain of the proposed AMAF-RIS structure grows linearly with the RIS aperture, in line with classical results for standard reflector antennas.  
\end{abstract}


\section{Introduction}  
\label{sec:intro}

\IEEEPARstart{W}{ireless} communication in the millimeter wave (mmWave) and sub-THz frequency bands has gained substantial attention in recent years, promising unprecedented data rates and ultra-low latency \cite{ted_thz}. 
It is by now well-understood that, at these high frequencies, classical wide-angle (e.g., cellular sectors) antennas and non-line-of-sight  (NLOS) rich scattering 
propagation, as in conventional sub-6GHz (i.e., the FR1 frequency tier in the 3GPP terminology \cite{38.101-1}) cannot offer acceptable range and coverage. 
In contrast, due to the small wavelength, very ``electrically'' large aperture antenna arrays can be implemented in a small form factor, offering 
the possibility of extremely high beamforming (BF) gains for directional line-of-sight (LOS) propagation. 
Applications may cover, for example,  wireless fronthaul, fixed point-to-multipoint wireless access (FWA), and multiuser MIMO through highly directional multiple 
beams per sector \cite{JSDM} in relatively small cells with LOS coverage (e.g., hotspots covering large indoor halls). 
It is also well understood that, for arrays with several hundreds (or thousands) of elements and signal bandwidth of 1GHz or more, 
fully digital processing (requiring a full RF chain per antenna element) is not feasible in terms of hardware complexity, A/D front-end throughput,  
and power consumption. Hence, the so-called hybrid digital-analog (HDA) approach is widely advocated, where the number of baseband antenna ports is much 
smaller than the number of array radiating elements, and the signal processing is split into an analog part in the RF domain, and a digital part in the baseband 
domain \cite{HDA1, HDA2, HDA3}. Among the many possible HDA architectures, the so-called 
one-stream per sub-array (OSPS), also referred to as ``partially connected'' architecture (e.g., see \cite{commit_bf1}) is particularly attractive since
the analog RF beamforming network reduces to stacking ``parallel'' independent modules, each implementing a single beam-steering array. 

Nevertheless, even for these simple HDA architectures, the complexity and power efficiency of large beam-steering active arrays is still
quite problematic. Hence, to realize the full potential of these frequency bands, innovative antenna configurations are called for. 
{\em Reflectarrays} have been known for a long time \cite{1963, Pozar_1997, 2003_steer_flat_ref}. More recently, reflective passive surfaces with electronically programmable phase shifts have been ``rediscovered'' under the name of Reflective Intelligent Surfaces (RIS). 
In the current research literature, RIS have been studied mainly as a way to modify the wireless multipath channel when the RIS is the far field of both the transmitter and the receiver \cite{RenzoMisLead1, Renzo1, EmilChannelEstimation, Emil_RIS_SP}. However, this approach is problematic since the reflected signal power is very weak in this case unless the physical size of the RIS is  enormous and impractical, e.g., as shown in \cite[(22)]{fau2}. 
Consider for example a transmitter-receiver (Tx-Rx), Tx-RIS, and RIS-Rx distances of 100m, 50m, and 50m, respectively. A RIS needs 3,300 (@ 5 GHz carrier) and 6,600 (@ 10 GHz carrier) elements in order to provide a link strength 
comparable to the direct LOS link. Taking the 5 GHz case as an example, 
let us consider a $58 \times 58$ standard rectangular array (SRA)  \cite[p. 236]{trees} RIS with 3,364 elements, the largest linear dimension is 
$1.414 \times 60 \lambda/2 \approx 82\lambda/2$.  The propagation time difference $\Delta \tau$ across the RIS of a planar 
wave impinging on the RIS at an angle $\psi$ with respect to its boresight direction is given by  
\begin{equation}
\label{eq:tau}
\Delta \tau = \frac{\text{sin}(\psi)\lambda D}{2c},
\end{equation}
where $c$ is the speed of light and $D$ denotes the largest linear dimension (the rectangle diagonal) of the RIS SRA normalized by $\lambda/2$. 
In this example, we have $\Delta \tau = 7.1~\text{ns}$. This means that in order for the classical ``narrowband'' channel model (widely used in the RIS literature) to hold,  the channel bandwidth $W$ should be limited such that the product $W \Delta \tau$ is sufficiently small. Imposing $W \Delta \tau \leq 0.125$ (\cite[p. 34, (2.47)]{trees}, yields  $W \leq 17.6~\text{MHz}$, which is clearly very limited for any reasonable 
application (only $0.35\%$ of the carrier frequency). Repeating this calculation for a 100 GHz carrier frequency and the same far-field Tx-RIS-Rx set-up as above, a  $183 \times 183$ SRA RIS is needed, yielding a bandwidth constraint of $W \leq 0.11 ~\text{GHz}$ (only $0.01\%$ of the carrier frequency). 
On the other hand, if the narrowband condition does not hold, then the widely used frequency-flat model of the Tx-RIS-Rx channel
of the type $\Hm_1 \Thetam \Hm_2$ where $\Hm_1$ and $\Hm_2$ are the complex baseband MIMO channel matrices between the Tx and the RIS, and between the RIS and the Rx, respectively, and $\Thetam$ is the diagonal matrix with unit-modulus elements, modeling the (frequency independent) 
RIS phase shifts, does not hold any longer. In this case, $\Hm_1$ and $\Hm_2$ are frequency-dependent (e.g., dependent of the subcarrier index of an OFDM modulation) and the optimization of the frequency-independent RIS phase shifts to achieve a desired 
communication performance is much more involved than in the frequency-flat case.\footnote{Notice that most of the early RIS literature has considered
this type of models with $\Hm_1$ and $\Hm_2$ frequency-independent rich scattering matrices with Gaussian zero-mean elements \cite{HDH1, HDH2, HDH3}.} 

\begin{figure}[h!]
\centerline{\includegraphics[width=8cm]{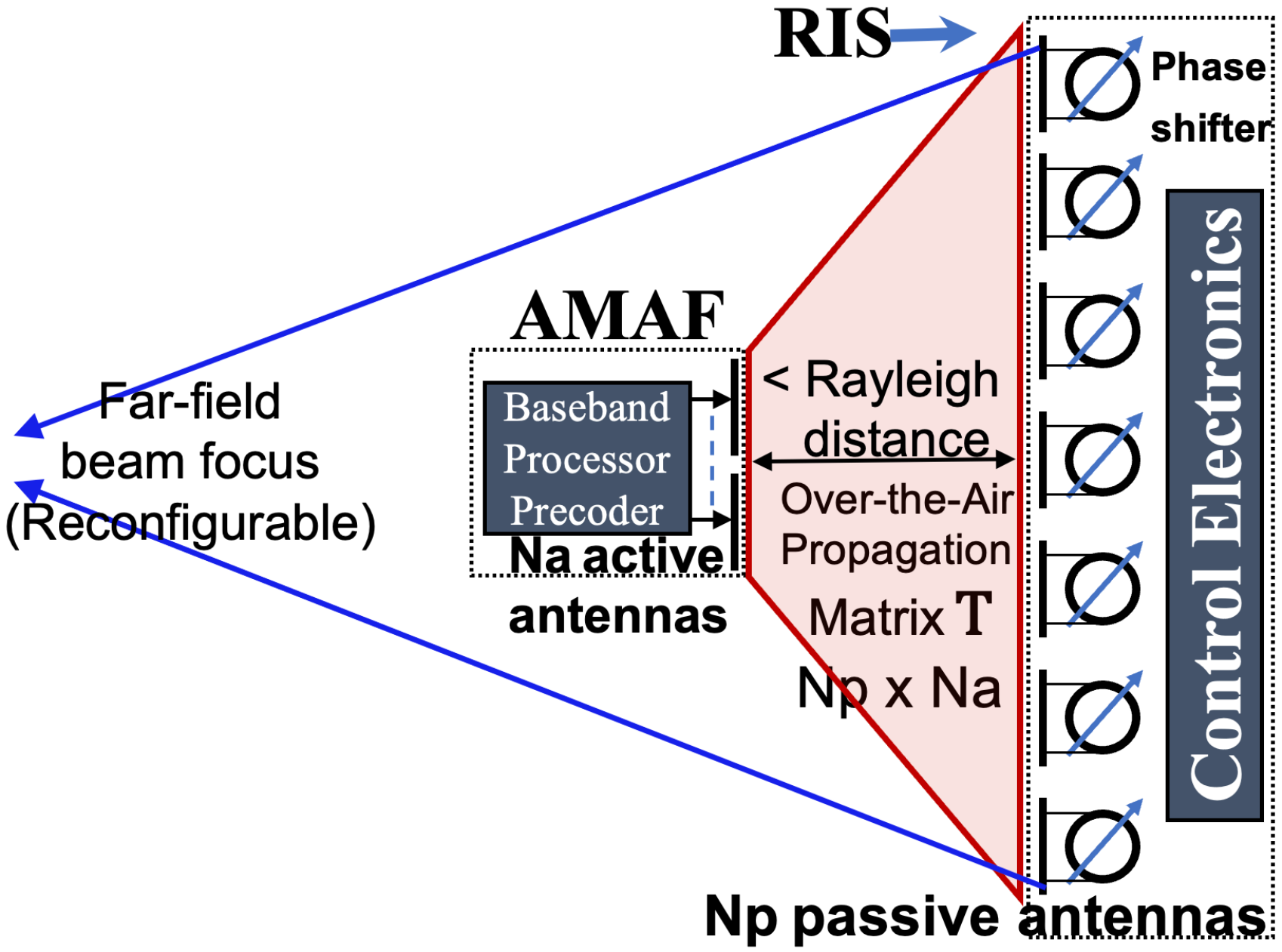}}
\caption{A RIS fed by an AMAF placed in its near field \cite{ICC2023}.}
\label{fig:sysmo}
\end{figure}

In line with \cite{RISBS}, \cite{buzzi}, \cite[Fig.~1(e)]{FAU_NF_RIS}, and \cite{2018MRA}, we propose the use of RIS in a reflectarray configuration where the RIS is in the near field of a small active multiantenna feeder (AMAF). In particular, in this paper we consider the downlink case, where the 
AMAF provides the data signals and the Tx power, and the user receivers are in the far-field of the RIS. 
In \cite{ICC2023}, we presented results on the beamforming and power efficiency properties of the AMAF-RIS module in a
2D geometry where both AMAF and RIS are standard linear arrays  and the AMAF is placed in its near field of the RIS 
with a focal length to diameter ($F/D$) ratio of 0.5 to 0.7, where $D$ is the size of the RIS~\footnote{In this paper, all lengths are normalized by $\lambda/2$, unless specified otherwise.} 
and the focal length $F$ is the AMAF-RIS distance. 
For the reader's convenience, we reproduce Fig. 1 of \cite{VTC2022, ICC2023} as Fig. \ref{fig:sysmo} here. 
We refer to our proposal as a ``new old idea" because reflectors in the near-field have been used for decades \cite{MarsCubeSat, TICRA, Yang}. Nevertheless,  our work is novel in the sense that the classical feed horn is replaced by the AMAF which allows a more refined design optimization, as it will be seen in the following. 

Our approach is motivated by the fact that, at very high carrier frequencies, power efficiency and hardware complexity are the main hurdles to be addressed, rather than spectral efficiency. It is well-known (see  \cite[Fig.~24]{Rui_Tut}) that  the RIS gain is 
maximized when the RIS is placed near to the transmitter or near to the receiver. In our case, this means that placing the RIS in the near-field of 
the AMAF maximizes energy efficiency. In addition, it is clear that using only small active arrays yields significantly reduced hardware complexity with respect to the hardware required by large active arrays.  Specifically, the contributions of this paper are:
\begin{itemize}
     \item We develop the full 3D beamforming design of the proposed AMAF-RIS architecture based on the singular value decomposition of the 
     near field channel matrix between the AMAF and the RIS for planar array configurations and provide results for a typical configuration for outdoor 
     small cell communications with LOS propagation.
     \item Multiuser, multibeam performance in terms of communication rate cumulative distribution functions (CDFs) is presented.
     \item We also study the impact of the RIS phase-shifter quantization and beam pointing errors. 
     \item We provide an accurate link budget analysis and a power efficiency analysis of the proposed design and compare with the power required by more 
     conventional architectures based on large active antenna arrays, in view of the state-of-the-art semiconductor technology \cite{ETH_PA_Survey}.
     \item In the appendix, we derive the linear scaling of the RIS gain with its size.
     Interestingly, this behavior is compliant with classical results for standard reflector antennas and it is different from 
     the quadratic scaling of the RIS-induced path gain observed when the RIS is in the far field of both the transmitter (feeder) and the receiver \cite{wu2019intelligent}. 
     As explained in \cite{Emil_RIS_SP}, though, the quadratic scaling is offset by the pathloss of the feeder-to-RIS channel, such that 
     eventually (and consistently with \cite{fau2}) the proposed architecture, where the feeder is in the near-field of the RIS, is much more power efficient. 
\end{itemize}

The rest of the paper is organized as follows: Section \ref{sec:3DModel} details the 3D geometry and system model. 
In Section \ref{sec:designs}, we present the AMAF-RIS principal eigenmode beamforming design, 
and consider the CDF of the achievable rate in LOS conditions. In Section \ref{sec:PE}, we provide the power efficiency analysis. Finally, 
Section \ref{sec:CONC} concludes the paper by summarizing the key highlights and indicating future research directions. 
 
{\bf Notations:} We shall use the following mathematical notations: sets by calligraphic letters $\mathbb{X}$. The sets of all real and all complex numbers are denoted as $\mathbb{R}$ and $\mathbb{C}$, respectively. Symbol $\text{x}^*$ denotes conjugate of a complex scalar $\text{x}$, $\|\xv\|$ is the Euclidean norm of a vector $\xv$, $[\cdot]^{\circ2}$ denotes Hadamard (element-wise) square, $[\cdot]^\transp$ is transpose, $[\cdot]^\herm$ is the Hermitian transpose, $|\xv|$ is a vector which contains the magnitudes of the elements in $\xv$, $|\Xm|$ is a matrix which contains the magnitudes of the elements in $\Xm$, and $\text{diag}(\xv)$ denotes a square diagonal matrix whose diagonal elements are given by the vector $\xv$.

\section{3D Geometry and System Model} 
\label{sec:3DModel}



\begin{figure}[h!] 
\centerline{\includegraphics[width=8cm]{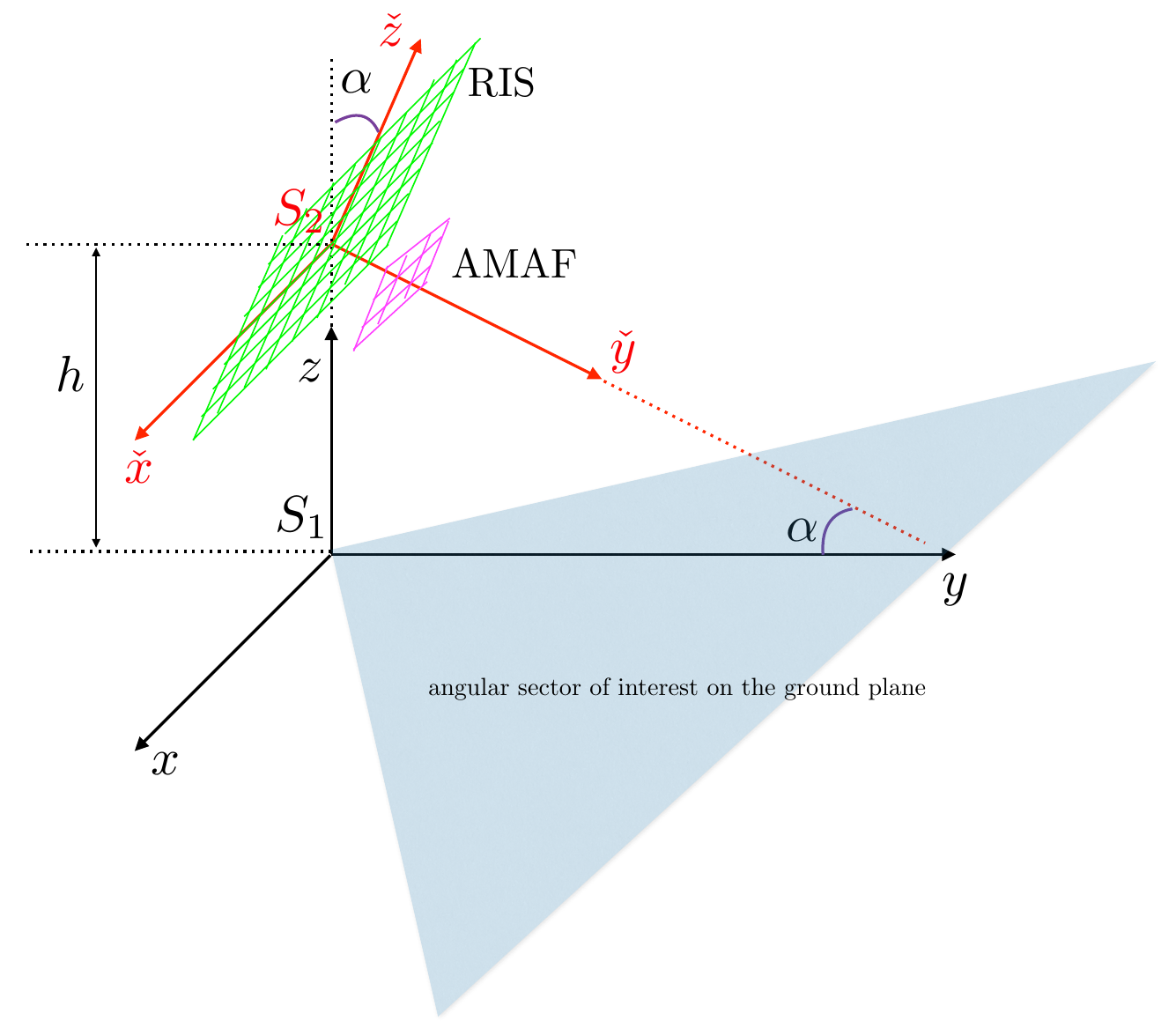}}
\caption{3D geometry.}
\label{3D-fig11}
\end{figure}

Consider the geometry in Fig.~\ref{3D-fig11} (distances are normalized by $\lambda/2$). 
We define two coordinate systems. S1 has its origin on the ground plane x-y. 
S2 has its origin in the center of the RIS, positioned at $(0,0,h)$ with respect to the S1 system, 
and is tilted by a rotation of $-\alpha$ in the z-y plane such that a rectangular planar panel centered at the origin of S2 in the x-z plane of S2 ``faces down'' 
toward the plane x-y of S1. This ``mechanical tilting'' is widely used for tower-mounted base stations to illuminate a certain region 
of the ground plane. In particular, we are interested in illuminating an angular sector in the ground plane of S1 as indicated in Fig.~\ref{3D-fig11}.

Let $\iv = (1,0,0)^\transp,\jv = (0,1,0)^\transp, \kv = (0,0,1)^\transp$ denote the three versors of S1 in the coordinate system S1, and 
$\check{\iv} = (1,0,0)^\transp, \check{\jv} = (0,1,0)^\transp, \check{\kv} = (0,0,1)^\transp$ denote the three versors of S2 in the coordinate system S2. 
The rotation matrix $\Rm_{12}$ of the transformation S1 $\mapsto$ S2 is obtained by expressing as columns the versors
$\iv,\jv,\kv$ in the rotated coordinates of S2. This yields
\begin{equation} 
\Rm_{12}  = \left [ \begin{array}{ccc}
1 & 0 & 0 \\
0 & \cos(\alpha) & -\sin(\alpha) \\
0 & \sin(\alpha) & \cos(\alpha) \end{array} \right ]. 
\end{equation}
It follows that a point $\pv = p_x \iv + p_y \jv + p_z \kv$ expressed in Cartesian coordinates with respect to S1 
can be expressed in Cartesian coordinates $\check{\pv} = \check{p}_x \check{\iv} + \check{p}_y \check{\jv} + \check{p}_z \check{\kv}$ 
with respect to S2 by applying a translation by $(0,0,h)$ and the above rotation, i.e., 
\begin{equation}
\check{\pv} = \Rm_{12} ( \pv - (0,0,h)^\transp).   \label{coord-transf}
\end{equation}

\begin{figure}[h!] 
\centerline{\includegraphics[width=6cm]{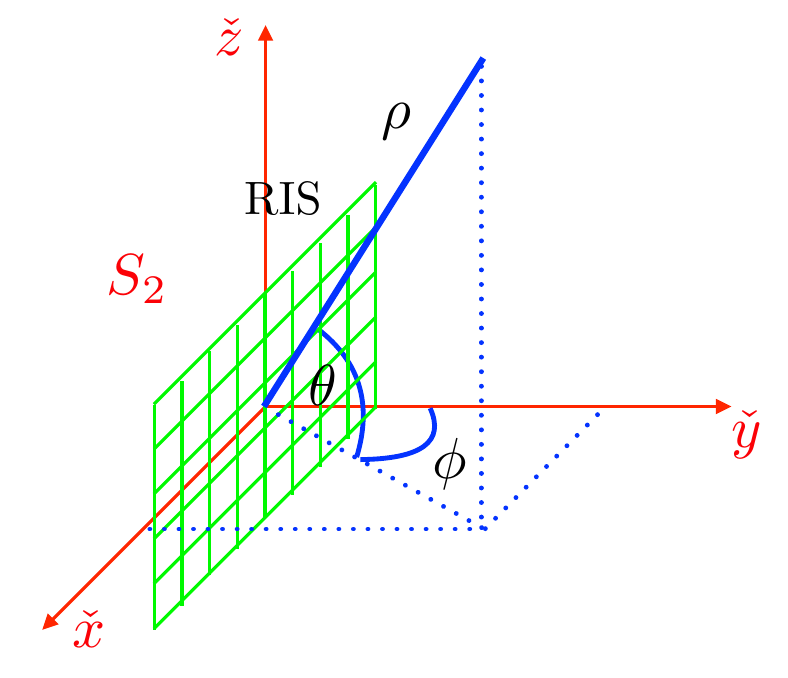}}
\caption{Spherical coordinates definition for the RIS coordinate system S2.}
\label{polarcoord1}
\end{figure}

It is also useful to define the spherical coordinates for S2 as given in Fig.~\ref{polarcoord1}. In particular, we define the azimuth angle $\phi$ with respect to the y-axis of S2 with the positive direction of the clockwise direction and the elevation angle $\theta$ with respect to the x-y plane of S2 with the positive direction ``up'' (toward the z-axis). 
In this way, the RIS boresight direction (normal to the RIS panel in the x-z plane of S2) 
corresponds to the angles $\phi = 0, \theta = 0$, and this is a line tilted down and impinging on the ground plane (xy plane of S1) 
by an angle $\alpha$ (see Fig.~\ref{3D-fig11}). 
%
Hence, a point of $\check{\pv}$ in S2 with spherical coordinates $(\rho, \phi, \theta)$ has
\begin{subequations}  \label{polar}
\begin{eqnarray}
\check{p}_x & = & \rho \sin \phi \cos \theta \\
\check{p}_y & = & \rho \cos \phi \cos \theta \\
\check{p}_z & = & \rho \sin \theta 
\end{eqnarray} 
\end{subequations}
By combining \eqref{coord-transf} and \eqref{polar} one can map any point on the ground plane of S1 to a far-field direction $(\phi,\theta)$ and a range $\rho$ with respect to the RIS. 


\subsection{RIS Array and Far-Field Array Response}
\label{subsec:array}

We consider a SRA RIS with reflecting elements equally spaced ($\lambda/2$ spacing) in a rectangular region in 
x-z plane of the system S2. The coordinates of the reflecting elements are given by 
\begin{equation}
\check{\pv}_{n,m} = \frac{1}{2} ( 2n - N_x + 1, 0, 2m - N_z + 1), 
\end{equation}
for $n \in \{0,1,\ldots, N_x-1\}$ and $m \in \{0,1,\ldots, N_z-1\}$, for a total of $N_p = N_x N_z$ elements.  
A planar wavefront impinging on the RIS panel at an angle $(\phi, \theta)$ has normal vector given by 
\begin{equation}
\nv(\phi, \theta) = (\sin \phi \cos \theta, \cos \phi \cos\theta,\sin\theta)^\transp.
\end{equation}
The phase shift (complex phasor term) of the $(n,m)$ element of the RIS is given by 
    \begin{align}
    a_{n,m}(\phi,\theta)
    =  & \exp \left ( - j \frac{2\pi}{\lambda} \frac{\lambda}{2} \check{\pv}_{n,m} \cdot \nv(\phi,\theta) \right )  \nonumber \\
    =  & \exp \big ( - j \frac{\pi}{2} ( (2n - N_x + 1) \sin\phi \cos \theta   \nonumber \\
    & + (2 m - N_z + 1) \sin\theta) \big ). 
    \label{eq:array-vec}
    \end{align}
%

We assume that each individual RIS element is a small patch antenna with antenna pattern $G_{\rm ris}(\phi,\theta)$.
In particular, for squared patch antennas, a commonly used model is the so-called axisymmetric cosine pattern (e.g., see \cite[(14)]{FAU_NF_RIS}, \cite[(17)]{Pozar_1997}, \cite[(2-31)]{Balanis_antenna_theo}) given by  $G_{\rm patch}(\theta, \phi) = 4 \cos^2(\psi)$,  
where $\psi$ is the angle of the direction $(\phi,\theta)$ with respect to the patch broadside, i.e., the y-axis of S2. 
It is immediate to see that  $\cos(\psi)$ is simply given by inner product of the normal vector $\nv(\phi,\theta)$ 
with the versor $\check{\jv}$ of the y-axis of S2 such that 
\begin{equation} \label{eq:psi}
\cos(\psi) = \nv(\phi,\theta) \cdot (0,1,0)^\transp = \cos \phi \cos \theta.
\end{equation}
It follows that
\begin{equation} \label{eq:patch}
G_{\rm patch}(\phi,\theta) = 4 \left ( \cos \phi \cos \theta \right )^2. 
\end{equation}
Notice that $G_{\rm patch}(\phi,\theta)$ has a half power beam width (HPBW) of $90 \degree$ and the power gain of 6 dBi\footnote{``dBi'' refers to the gain of a directive antenna element over a theoretical isotropic antenna element with the same radiation intensity in all directions.} which is typical of microstrip patch antennas widely used in RIS hardware implementation. Since the RIS needs to serve mobile UEs over a large beam scan space, a wide beam patch antenna is suitable for the RIS to minimize beam cusping losses \cite{Fall_VTC_2022, VTC2021}. 

Now, suppose that the RIS is excited by an impinging near-field signal (e.g., generated by the AMAF) 
such that the complex signal amplitude at each element is $u_{n,m}$, and suppose that the RIS can further impose a 
phase rotation $w_{n,m} = e^{j \mu_{n,m}}$ for each $n,m$ element. 
The resulting radiation pattern of the RIS with feed-induced complex amplitude profile $u_{n,m}$ and phase shift profile $w_{n,m}$ 
as  a function of the angle direction $(\phi,\theta)$ is given by 
\begin{equation} 
G(\phi,\theta) = G_{\rm patch}(\phi,\theta) \left | \sum_{n=0}^{N_x-1} \sum_{m = 0}^{N_z - 1} w_{n,m} u_{n,m} a_{n,m}^*(\phi,\theta) \right |^2.
\end{equation}
For convenience, we define the ``tapered'' array response coefficients as
    \begin{equation}
    \tilde{a}_{n,m}(\phi,\theta) = u_{n,m} a_{n,m}(\phi,\theta).
    \label{eq:def1}
    \end{equation} 
Collecting $\{\tilde{a}_{n,m}(\phi,\theta)\}$ and $\{w_{n,m}\}$ into two $N_p \times 1$ vectors $\tilde{\av}(\phi,\theta)$ and $\wv$, the 
RIS radiation pattern can be compactly written as
\begin{equation} 
G(\phi,\theta) = 4 \left ( \cos \phi \cos \theta \right )^2 \left | \tilde{\av}(\phi,\theta)^\herm \wv \right |^2.\label{eq:bfpattern}
\end{equation}
We define the RIS gain $\Gamma$ as the maximum possible value of $G(\phi,\theta)$ from (\ref{eq:bfpattern}). This is obtained
for $\phi = \theta = 0$, letting $\wv = \onev$ (the all-one vector), and yields
\begin{equation} 
\Gamma = 4 \left | \sum_{n=0}^{N_x-1} \sum_{m = 0}^{N_z - 1} u_{n,m} \right |^2. \label{eq:gamma}
\end{equation}

In \cite{2013array}, it has been shown that RIS radiation performance characterization using array theory, as in (\ref{eq:bfpattern}), closely matches the results from aperture theory, where full-wave simulations have shown that both methods can be used equivalently for a reliable calculation of the general pattern shape, main beam direction, beamwidth, and sidelobe levels. Therefore, we choose the array theory based RIS radiation pattern calculation, which is faster and requires less computational effort.


\subsection{AMAF-RIS Near-Field Propagation Matrix}
\label{subsec:T}

We consider the AMAF formed by a SRA of size $N_h \times N_v$ placed at distance $F$ from the RIS \cite[Table I]{ICC2023}. The number of active antenna elements is therefore $N_a = N_h N_v$.  
For the {\em narrowband model} to hold, it is necessary that the maximum propagation time difference $\Delta \tau$ between any two paths across the AMAF-RIS structure is significantly less than the reciprocal of the signal bandwidth $1/W$. We have already seen that the propagation time difference across the RIS for 
a planar wavefront impinging at angle $\psi$, with respect to the RIS boresight direction, is given by \eqref{eq:tau} with $D = \sqrt{N_x^2 + N_z^2}$. 
In addition, we need to consider the propagation time difference between any two pairs of RIS and AMAF elements. 
This is upper bounded by the maximum distance difference, divided by $c$.

\begin{figure}[h!] 
\centerline{\includegraphics[width=6cm]{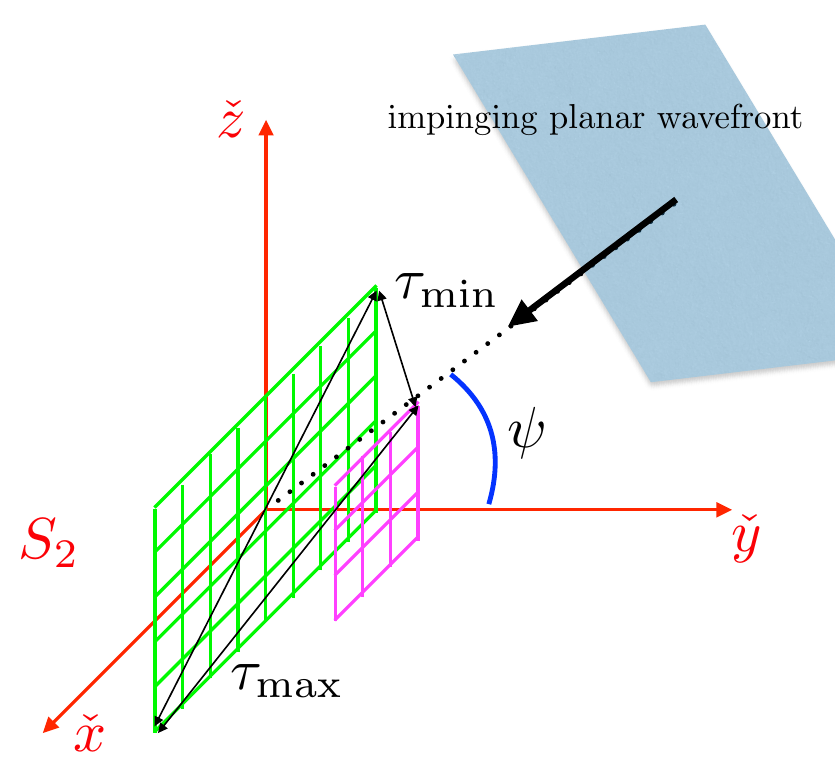}}
\caption{Maximum and minimum path delay for a planar wave propagation across the AMAF-RIS structure.}
\label{fig:maxmin-delay}
\end{figure}

As shown in Fig. \ref{fig:maxmin-delay}, the maximum distance is given by 
taking two opposite corner elements, e.g., the element $\check{\pv}_{0,0}$ of the RIS, and the element 
$\check{\qv}_{N_h-1,N_v-1}$ of the AMAF, were the AMAF elements have coordinates
\begin{equation}
\check{\qv}_{g,k} = \frac{1}{2} ( 2g - N_h + 1, F, 2k - N_v + 1)^\transp, 
\end{equation}
for $g \in \{0,1,\ldots, N_h-1\}$, $k \in \{0,1,\ldots, N_v-1\}$. 
It follows that the maximum propagation time
\begin{equation} \label{eq:MaxDelay}
\tau_{\rm max} = \frac{\lambda}{2c} \left ( \sin(\psi) \sqrt{N_x^2 + N_z^2} + \| \check{\pv}_{0,0}-\check{\qv}_{N_h-1,N_v-1}\| \right ),
\end{equation}
and the minimum propagation time
\begin{equation} \label{eq:MinDelay}
\tau_{\rm min} = \frac{\lambda}{2c}  \|\check{\pv}_{0,0}-\check{\qv}_{0,0}\| .
\end{equation}
Thus, the maximum time difference is upper bounded by 
    \begin{align}
    \Delta \tau_{\max}
    \!\!=  & \tau_{\rm max} - \tau_{\rm min}  
    \nonumber \\
    \!\!=  & \frac{\lambda}{2c} \left ( \sin(\psi) \sqrt{N_x^2 + N_z^2} + \| \check{\pv}_{0,0}-\check{\qv}_{N_h-1,N_v-1}\| \right )   \nonumber \\
    & - \frac{\lambda}{2c}  \|\check{\pv}_{0,0}-\check{\qv}_{0,0}\| .
    \label{eq:max_delay_spread}
    \end{align}
    
In the example of Section \ref{sec:designs}, we consider 
$N_x = N_z = 16$, $N_h = N_v = 4$, $F = 8$, and the maximum angle of the coverage region 
$\psi = 63.3\degree$ (from \eqref{eq:psi}, $\psi= \cos^{-1}~(\cos \phi \cos \theta)$, the cell azimuth $\phi=+/- 60\degree$, and elevation $\theta=+/- 26.07\degree$). With these parameters, we have $\Delta \tau_{\max} = 117.9 ~{\rm ps}$. Imposing again 
the constraint $W \Delta \tau_{\max} \leq 0.125$, we find that the narrowband model holds for signal bandwidths up to $W \leq 1.06 ~{\rm GHz}$. Larger bandwidths result in inter-symbol interference (ISI). For example, for a system bandwidth $W = 5~{\rm GHz}$, the resulting 
inter-symbol interference (ISI) length is 
\begin{equation} \label{eq:ISI_length}
\Lambda = \lceil \Delta \tau_{\max} W \rceil = \lceil 0.6 \rceil = 1
\end{equation}
``chip'' (time domain symbol). This means that an orthogonal frequency division multiplexing (OFDM) system with cyclic prefix (CP) = 1 (or single-carrier system with CP and frequency domain equalization) is sufficient to eliminate the ISI. In turn, a very moderate block length $N_{\rm F}$ (e.g., 8 subcarriers) is already sufficient to yield a low CP precoding overhead.   


For convenience, we enumerate the RIS and AMAF elements row by row (or in any suitable order, as long it is consistent) using indices 
$k \in \{0, \ldots, N_p -1\}$ and $\ell \in \{0,\ldots, N_a - 1\}$, respectively. Letting $r_{k,\ell}$ denote the distance between 
the $k$-th RIS element and the $\ell$-th AMAF element, and letting $(\varphi_{k,\ell},\vartheta_{k,\ell})$ the angle at which they see each other with respect to their own
normal (boresight) direction, the near-field wideband propagation matrix $\Tm \in \CC^{N_p \times N_a}$ between the $N_a$ AMAF elements and the $N_p$ RIS elements
has entries 
\begin{equation} \label{eq:T}
T_{k,\ell}(f)  = \frac{\sqrt{E_A(\varphi_{k,\ell},\vartheta_{k,\ell})E_R(\varphi_{k,\ell},\vartheta_{k,\ell}})}{2\pi r_{k,\ell}}~ e^{ - j\pi (r_{k,\ell}+2f \tau_{k,\ell})},
\end{equation}
where $E_A(\varphi,\vartheta)$, $E_R(\varphi,\vartheta)$ are the AMAF and RIS element radiation patterns, respectively, $\tau_{k,\ell}  = \lambda_0 r_{k,\ell}/ (2c) = r_{k,\ell}/(2 f_0)$, $f_0$ is the carrier frequency, and $f \in [-W/2, W/2]$. 
Note that for centered AMAF and RIS (see Fig.~\ref{3D-fig11}), the RIS element and the AMAF element see each other at corresponding 
opposite angles. In this work we have considered the same patch element with the axisymmetric pattern 
such that $E_A = E_R = G_{\rm patch}$ given in  (\ref{eq:patch}), although the approach can be generalized to particular designs especially 
for the AMAF radiating elements. 
While the AMAF is in the near field of the RIS, the individual antenna elements of the AMAF and the RIS are in each other's far field. In fact, 
(\ref{eq:T}) is the Friis transmission equation \cite[eq. (2-119)]{Balanis_antenna_theo} in the magnitude form, along with the inclusion of the distance dependent and the (baseband) subcarrier dependent phase terms.  

For an $N_{\rm F}$ subcarriers OFDM system, we let 
\begin{equation} \label{eq:FST}
\Tm[\nu] = \left [ T_{k,\ell}(f_\nu) \right ],
\end{equation}
denote the AMAF-RIS matrix at subcarrier index $\nu = 0, \ldots, N_{\rm F}-1$, where 
$f_\nu = - W/2 + \nu \Delta f$ and $\Delta f = W/N_{\rm F}$ is the subcarrier spacing. 
The resulting CP overhead is  $\Lambda/(\Lambda+N_{\rm F})$. In our running example, with $\Lambda = 1$ and $N_{\rm F} = 8$ this yields a very 
moderate redundancy of just $1/9$.  


\subsection{AMAF (Pragmatic) Precoding Design}
\label{subsec:EModes}

We consider the singular value decomposition (SVD) of $\Tm(0)$ (the near-field matrix at $f = 0$, i.e., center-bandwidth)
\begin{eqnarray} \label{eq:SVD}
\Tm(0)  = \Um \Sigmam \Vm^\herm = \sum_{\ell = 1}^{N_a} \sigma_\ell \uv_\ell \vv_\ell^\herm.
\end{eqnarray} 
The AMAF is formed by a (small) array of active elements, each of which is driven by a complex vector modulator, i.e., 
it is controlled in both amplitude and phase. Since the columns $\{\vv_\ell\}$ of $\Vm$ form a unitary basis for $\CC^{N_a}$, 
any precoding vector $\bv \in \CC^{N_a}$ at the AMAF can be written as
$\bv = \sum_{\ell=1}^{N_a} \mu_\ell \vv_\ell$ where, 
without loss of generality, $\sum_{\ell=1}^{N_a}  |\mu_\ell|^2 = 1$ is imposed for transmit power normalization.
For a given choice of $\bv$, the resulting complex amplitude profile on the RIS elements at subcarrier $\nu$ is given by 
\begin{equation} 
\uv[\nu] = \Tm[\nu] \bv.   \label{eq:risuca}
\end{equation}
The AMAF-RIS array gain $G(\phi,\theta)$ defined in \eqref{eq:bfpattern} is in general dependent on the subcarrier index. 
In particular, we define $G_\nu(\phi,\theta)$ as given by \eqref{eq:bfpattern} after replacing $u_{n,m}$ with 
$u_{n,m}[\nu]$ where these are the component of $\uv[\nu]$ in \eqref{eq:risuca} after a suitable index re-arrangement.

Both the AMAF precoding vector $\bv$ and the RIS phase profile vector $\wv$ are frequency independent.
Therefore, we must choose $\bv, \wv$ in order to achieve a desired performance considering the whole system bandwidth, i.e., 
all the subcarriers. However, in this work we observe that, in the regime of system parameters considered in our examples, 
an excellent pragmatic choice consists of choosing $\bv = \vv_1$ (principal eigenmode of the AMAF-RIS 
transmission matrix at the center frequency) and impose a phase unwrapping at the RIS such that 
the components of $\uv = \Tm(0)\vv_1 = \sigma_1 \uv_1$ are real and positive. 
We refer to this choice as the  pragmatic {\em principal eigenmode} (PEM) design. 
This creates a template beam pattern given by  \eqref{eq:bfpattern} with an excellent footprint shape on the ground plane, centered at the RIS boresight 
direction $(\phi = 0, \theta = 0)$. It turns out that this ground footprint remains almost identical over the whole 
signal bandwidth, i.e., for all subcarrier indices $\nu$. This is due to the fact that the variations of the AMAF-RIS transmission matrix 
over the channel bandwidth is very moderate in the regime of interest (in fact, as seen before, 
the product $W \Delta \tau_{\max}$ is just equal to 0.6 in the case at hand). Of course, this pragmatic design choice may yield degradation of the beam pattern
at the edge of the channel bandwidth for larger $W$ and/or larger array dimensions, yielding a larger $\Delta \tau_{\max}$. 

With a slight abuse of notation, after incorporating the phases of $\uv_1$ into $\wv$,  the complex amplitude profile on the RIS elements at subcarrier index $\nu$ is given the  elementwise phase rotated vectors 
\begin{equation} 
\uv[\nu] = \left ( \Tm[\nu] \vv_1 \right ) \odot  \left [ \exp(-j \angle \uv_1) \right ],   \label{re:ew}
\end{equation}
where $\odot$ is Hadamard (elementwise) product, and $\left [ \exp(-j \angle \uv_1) \right ]$ is a vector of phase shifts such that 
$\uv_1 \odot \left [ \exp(-j \angle \uv_1) \right ] \in \RR_+^{N_p}$. 

Finally, beam steering of the template boresight beam to a desired direction $(\phi_0,\theta_0)$ 
is obtained by imposing a linear phase shift profile in the form $\wv = \av(\phi_0,\theta_0)$ on top of the phase unwrapping of $\left [ \exp(-j \angle \uv_1) \right ]$. 

For narrowband systems, where $\Tm = \Tm(0)$ is frequency-independent, PEM beamforming yields the maximum power transfer 
between the AMAF and the RIS for a given array geometry. 
Of course, in cases where the AMAF-RIS transmission matrix is more significantly frequency dependent,
the pragmatic PEM design does not achieve optimal power transfer on all subcarriers. 

\begin{rem} \label{rem1}
It is important to notice that the proposed PEM beamforming  needs no online optimization and the 
coefficients $\bv = \vv_1$ of the AMAF are fixed (i.e., they can be hard-wired and carefully calibrated in the production process). 
This is a very attractive feature of the proposed architecture.  \hfill $\lozenge$
\end{rem}

\subsection{Hybrid Beamforming Structure}  
\label{subsec:HybridBF}

\begin{figure}[h!]
\centerline{\includegraphics[width=7cm]{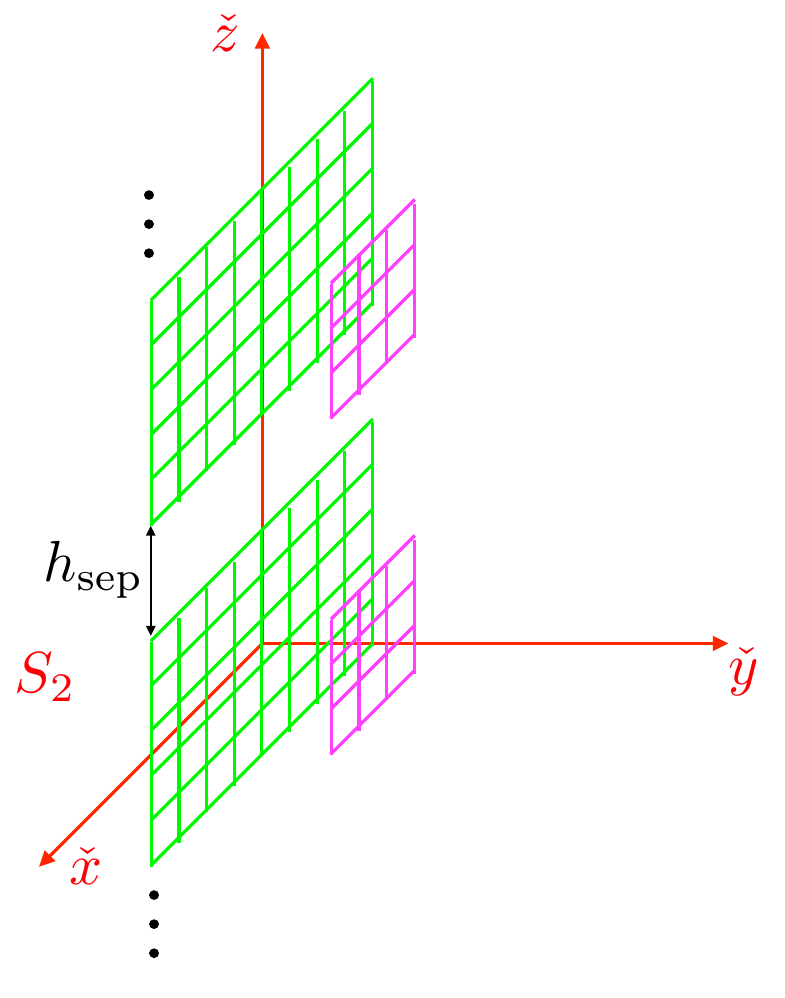}}
\caption{Stacked OSPS structure with over-the-air beamforming.}
\label{3D-fig3}
\end{figure}

In the spirit of OSPS multiuser MIMO design \cite{commit_bf1}, we consider the stacking 
of $K > 1$ AMAF-RIS basic modules with relatively small separation (e.g., the vertical stacking of Fig.~\ref{3D-fig3}). 
For the sake the far-field propagation, the modules can be considered as co-located. 
However, the distance is chosen large enough such that the mutual interference between the modules is very limited.  
From a mathematical viewpoint, any stacking yields the same structure with of course a different global matrix $\Tm[\nu]$.
In this case, the matrix $\Tm[\nu]$ (for every subcarrier $\nu$) has dimensions $K N_p \times K N_a$ and can be written as a 
$K \times K$ block matrix. Each block $\Tm_{i,j}[\nu]$ of dimensions $N_p \times N_a$ represents the propagation 
between the $j$-th AMAF array and the $i$-th RIS array at subcarrier $\nu$. We define also
\begin{equation} 
\Wm_i = \diag(\wv_i),
\end{equation}
as the diagonal $N_p \times N_p$ matrix of the phase shifts of RIS $i$, and $\bv_j$ to be the $N_a \times 1$ precoding vector of AMAF $j$. Hence, 
defining the block-diagonal matrices  $\Wm = \diag(\Wm_i : i = 1, \ldots, K)$ and $\Bm = \diag(\bv_j : j = 1, \ldots, K)$, the $\nu$-th subcarrier global 
transmission matrix of dimensions $K N_p \times K$ from the $K$ baseband antenna ports (each driving one AMAF) and the 
RIS reflecting elements is given by \eqref{eq:DTB2}.
\begin{align}
&  \Wm \Tm[\nu] \Bm  =  \nonumber \\
&   \left [ \begin{array}{cccc} 
\Wm_1 \Tm_{1,1}[\nu] \bv_1 & \Wm_1 \Tm_{1,2}[\nu] \bv_2 & \cdots & \Wm_1 \Tm_{1,K}[\nu] \bv_K \\
\Wm_2 \Tm_{2,1}[\nu] \bv_1 & \Wm_2 \Tm_{2,2}[\nu] \bv_2  & \cdots & \Wm_2 \Tm_{2,K}[\nu] \bv_K \\
\vdots &  & \ddots  & \vdots \\
\Wm_K \Tm_{K,1}[\nu] \bv_1 & \Wm_K \Tm_{K,2}[\nu] \bv_2 & \cdots & \Wm_K \Tm_{K,K}[\nu] \bv_K 
\end{array} \right ]  \label{eq:DTB2}
\end{align}

In this work we consider  identical  AMAF-RIS subarrays, such that we have $\Tm_{1,1}[\nu]= \cdots =\Tm_{K,K}[\nu]$. 
Our pragmatic design considers that the AMAF precoding vectors $\bv_j$ are designed as said before, based only on the 
principal eigenmode of the propagation matrix $\Tm_{j,j}(0)$ in isolation. Then, we have $\bv_j = \vv_1$ for all $j = 1, \ldots, K$. 
The cross-talk between the stacked arrays is controlled by setting the physical distance $h_{\rm sep}$ as shown in Fig.~\ref{3D-fig3}. 
It turns out that since the distance $F$ between each AMAF and its corresponding RIS is chosen such that 
most of the radiated RF power of the AMAF is actually transferred to the RIS (also due to the PEM precoding), 
even a moderate distance $h_{\rm sep}$ ensures negligible cross-talk between the subarrays. 

Consider now a multiuser MIMO scenario where a base station serves multiple users located in the coverage area on the ground plane (see Fig.~\ref{3D-fig11}).
Since the $K$-stacked structure can send up to $K$ downlink data streams,  the downlink 
scheduler  chooses groups of $K$ users to be served on the same time slot by spatial multiplexing. 
The resulting $K \times K$ LOS matrix (per subcarrier) between the $K$ base station antenna ports and the $K$ (far-field) users is given by 
\begin{equation}
\label{eq:BB_MIMO}
    \Hm[\nu] = \Am^\herm \Wm \Tm[\nu] \Bm   
\end{equation}
where 
\begin{equation}
\label{eq:Amat}
 \Am = 2 \left[\text{cos} \phi_1 \text{cos} \theta_1 \av(\phi_1,\theta_1), \ldots, \text{cos} \phi_K \text{cos} \theta_K \av(\phi_K,\theta_K) \right ] 
\end{equation}
is the $K P \times K$ array containing the steering vectors whose elements are given by (\ref{eq:array-vec}) for the stacked RIS array to the $K$ 
users, where each user $k$ is seen at an angle $(\phi_k,\theta_k)$ with respect to the coordinate system S2 of the RIS. Notice that, for simplicity, 
we neglect the subcarrier dependent (small) variations in $\Am$ in \eqref{eq:Amat} because the path delay difference from the RIS elements to any user in the far-field is negligible with respect to $1/W$.  Notice also that we include the factors $\sqrt{G_{\rm patch}(\phi_k,\theta_k)} = 2 \cos \phi_k \cos \theta_k$ 
weighting the array steering vectors  $\av(\phi_k,\theta_k)$ because the RIS gain in each given user direction must also take into account the directivity of the RIS elements. 
The $K \times K$ matrix $\Hm[\nu]$ in \eqref{eq:BB_MIMO} yields the baseband matrix of a ``low dimensional'' MU-MIMO channel 
with $K$ data streams and $K$ users, at subcarrier $\nu$. Ideally, we want the matrix $\Hm[\nu]$ to be strongly diagonal-dominant, which allows us to dispense 
with baseband signal processing techniques such as zero-forcing, etc., which are commonly used to tackle inter-stream interference 
in cases of non-diagonal $\Hm[\nu]$.

Assuming perfect timing and phase recovery at each user receiver, the achievable 
communication rate to the $k$-th user (BF only, BB precoding) under Gaussian single-user capacity achieving codebooks and treating multiuser interference as noise is given by  
\begin{equation}
\label{eq:Rate}
 R_k = \frac{1}{N_{\rm F}} \sum_{\nu=0}^{N_{\rm F}-1} 
 \log_2 \left(1+ {\rm SINR}_{k,\nu}\right),
\end{equation}
bits per complex signal dimension (equiv.,  bits/s/Hz),
where ${\rm SINR}_{k,\nu}$ is the signal to interference-plus-noise ratio over the subcarrier index $\nu$ for the $k$-th user,
    \begin{align}
    {\rm SINR}_{k,\nu} & = 
    \frac{| H_{k,k}[\nu]|^2 P_{\rm amaf}}{W N_0/L_k + \sum_{j=1, j \neq k}^{K}| H_{k,j}[\nu]|^2 P_{\rm amaf}}  
    \label{eq:SINR}
    \end{align}
where $H_{k,j}[\nu]$ is the $(k,j)$-th element of $\Hm[\nu]$, $P_{\rm amaf}$ is the total AMAF output RF power, 
$N_0$ is the complex baseband AWGN power spectral density, and $L_k = (\lambda / (4\pi d_k))^2$  is the free-space pathloss due to distance $d_k$, in meters, between user $k$ and the base station.  
Notice that for given angles in LOS condition the channel matrix $\Hm[\nu]$ is deterministic, and hence any standard synchronization
(carrier frequency, timing, and phase) at the user receivers can easily achieve (almost) ideal coherent detection. 
All the usual problems related to ``imperfect channel state information'' that appear in typical wireless communication scenarios do not play any significant 
role here. 

In (\ref{eq:Rate}) we have assumed that all the downlink data streams are transmitted at the same power $P_{\rm amaf}$. 
As it will be clear in the link-budget analysis of Subsection \ref{subsec:linkbudget}, the OSPS architecture is subject to a per-stream 
(or per antenna port) power constraint. Hence, it is not possible to optimize the power allocation of the total Tx power $K P_{\rm amaf}$ across 
the data streams.  Furthermore, as we shall see in the performance examples of Section \ref{sec:designs}, the matrices $\Hm[\nu]$ 
obtained with PEM precoding are strongly diagonally dominant. Since the multiuser interference is essentially negligible, in this case 
it does not make sense to attenuate the power of some data stream to mitigate interference to other streams. 
Indeed, for the example of this paper, the best choice consists of transmitting at maximum RF output power from all the AMAFs. 

\section{A Multiuser MIMO Example} 
\label{sec:designs}

\begin{figure}[h!] 
\centerline{\includegraphics[width=7cm]{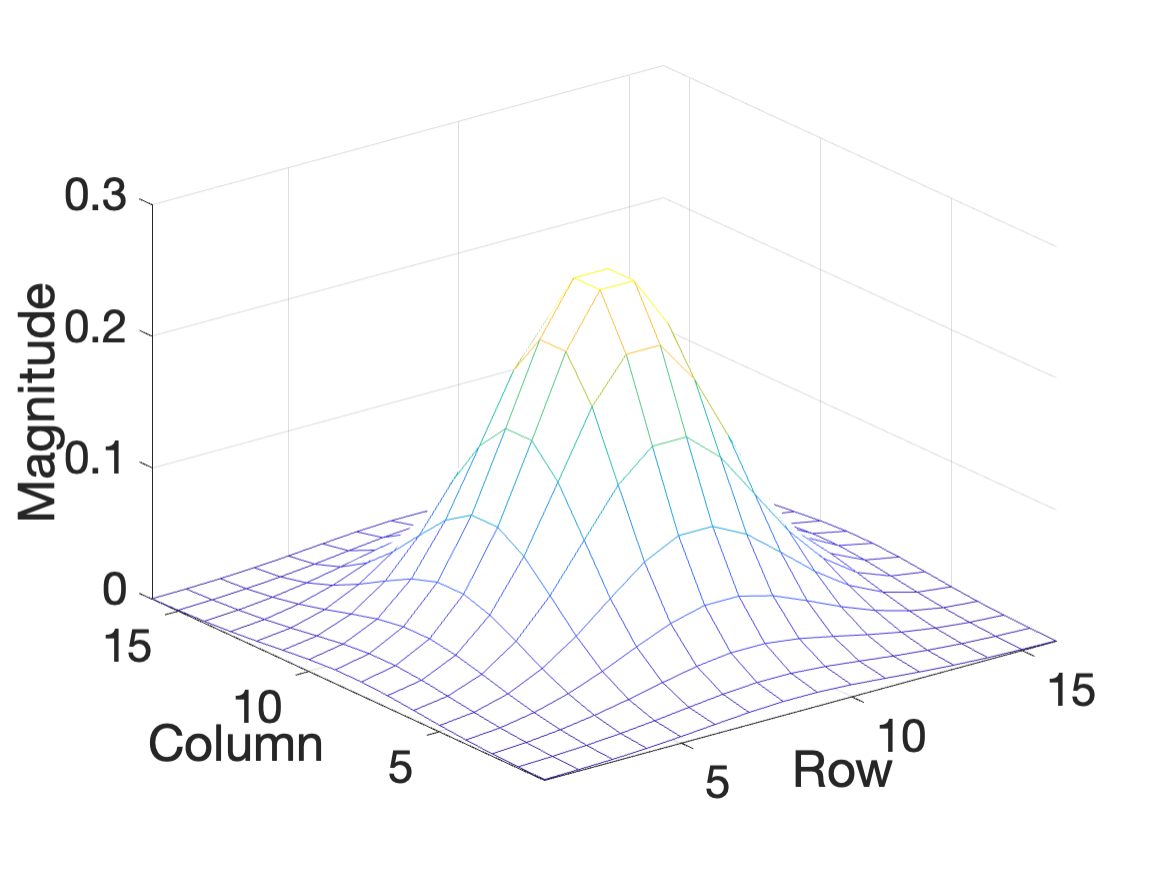}}
\caption{The $16 \times 16$ RIS PEM amplitude profile.}
\label{fig:RIS_Ex}
\end{figure}

\begin{figure}[h!] 
\centerline{\includegraphics[width=7cm]{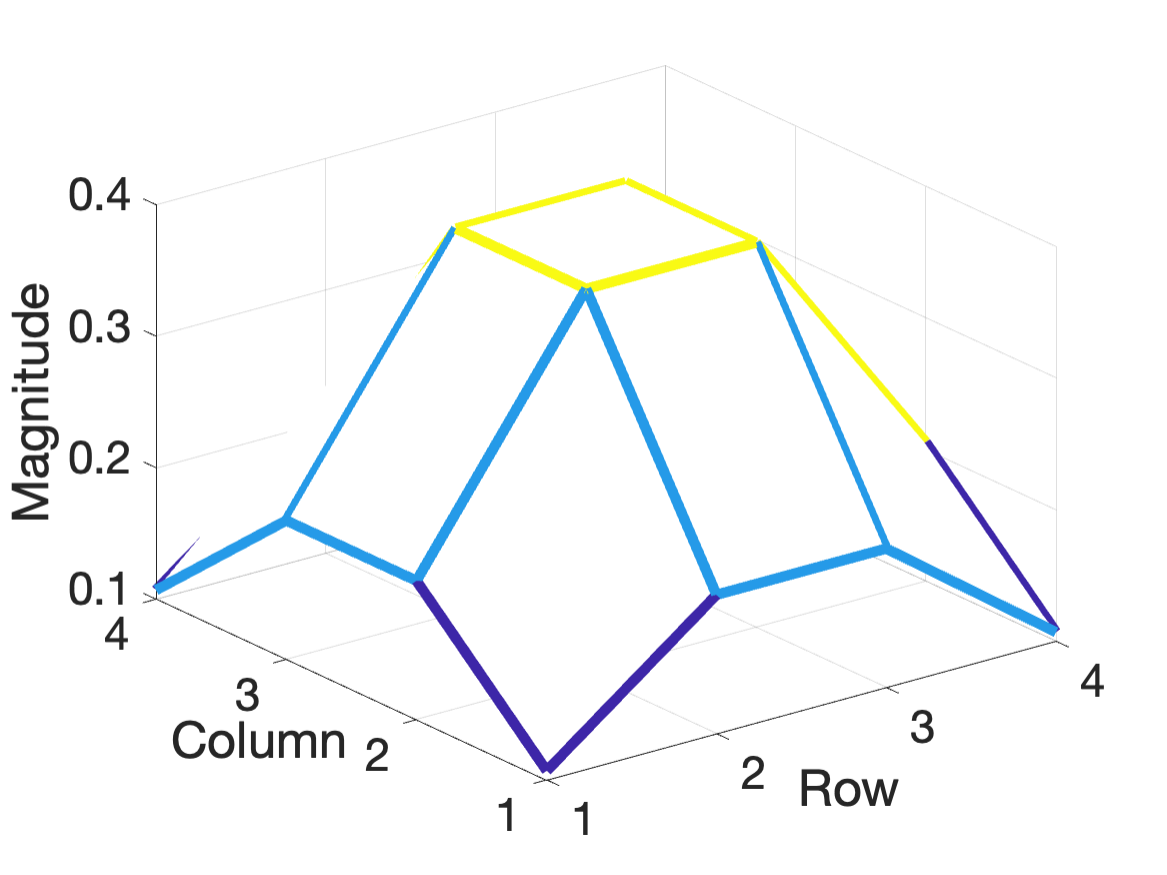}}
\caption{The $4 \times 4$ AMAF PEM amplitude profile.}
\label{fig:AMAF_Ex}
\end{figure}

We present a case study where the RIS and the AMAF are SRAs of size $N_x = N_z = 16$ and $N_h = N_v = 4$, respectively.  
Based on the observations in \cite[section III.A]{ICC2023}, we choose
a distance $F = 8$. This yields a good balance between the ``energy capture'' of the RIS with respect to the radiation pattern of the AMAF and the 
ability of the RIS to create a nicely shaped directional beam. Qualitatively, if $F$ is too small only a central portion of the RIS is illuminated by the AMAF, i.e., the elements of the RIS away from the center play no role in beamforming. If $F$ is too large, then a large fraction of the RF power radiated by the AMAF is lost in space since the solid angle covered by the RIS is too small. With an empirical tuning (this is just a one-dimensional line search), we find that 
in the case at hand $F = 8$ with the PEM beamforming yields a nice tapered amplitude profile on the RIS (magnitude of the components $\uv_1$, we denote as $|\uv_1|$), as shown in Fig.~\ref{fig:RIS_Ex}. 
Also noteworthy is the AMAF amplitude profile $|\vv_1|$ shown in Fig.~\ref{fig:AMAF_Ex}. 

We notice that PEM beamforming induces a symmetric amplitude profile at both the AMAF and the RIS. 
By placing the AMAF in the near field of the RIS, a very convenient ``taper'' of the RIS amplitude profile is induced. In turns, this yields very low side lobes of the radiation pattern (the same effect was observed in \cite{ICC2023} for the 2D geometry with linear arrays). 
In Fig.~\ref{fig:RIS_Ex} and Fig.~\ref{fig:AMAF_Ex}, the RIS  and AMAF tapers in $|\uv_1|^{\circ2}$ and $|\vv_1|^{\circ2}$
(i.e., the ratio between the  maximum and the minimum squared elements) are 58.9 dB and 11.3 dB, respectively. 

\begin{figure}[h!] 
\centerline{\includegraphics[width=8cm,height=4cm]{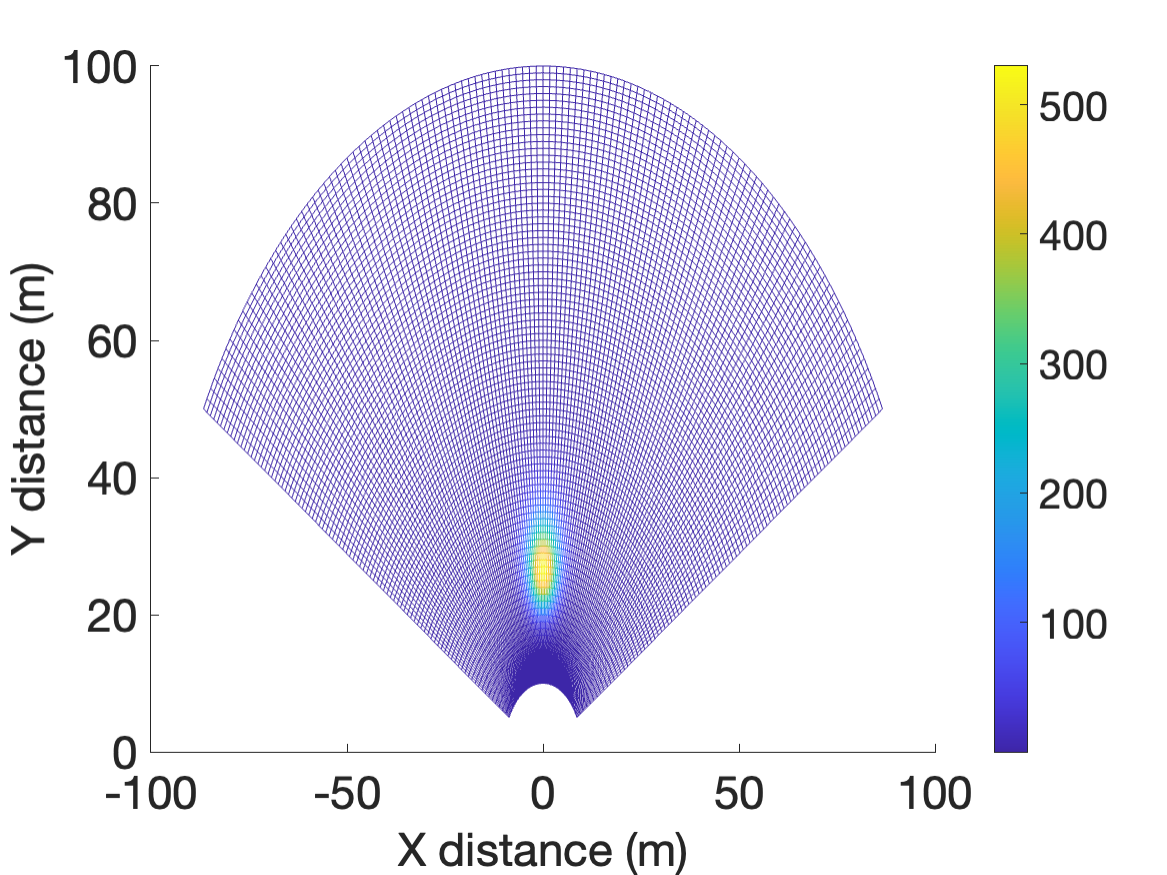}}
\caption{Ground footprint of the RIS PEM spot beam at the RIS S2 boresight, the center-bandwidth $f = 0$.}
\label{fig:CenterSpot0}
\end{figure}

\begin{figure}[h!] 
\centerline{\includegraphics[width=8cm,height=4cm]{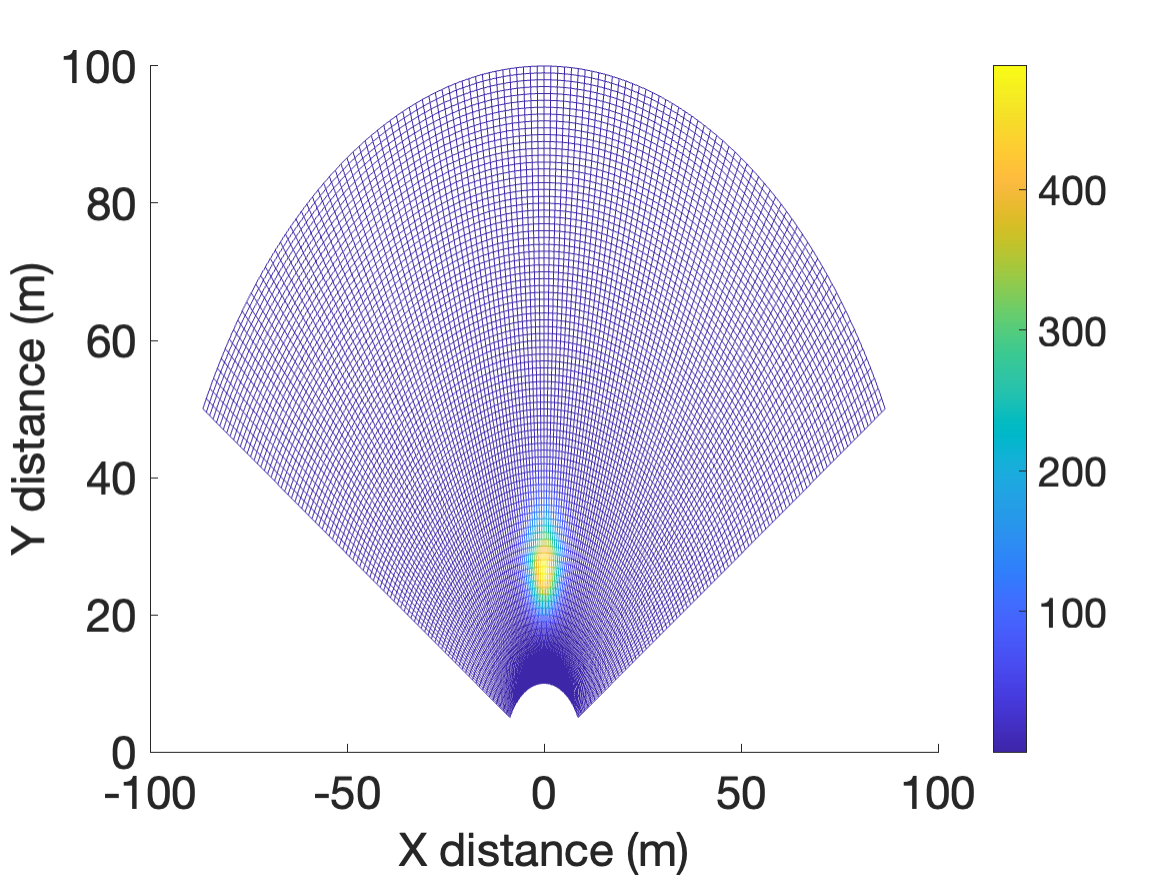}}
\caption{Ground footprint of the RIS PEM spot beam, the edge frequency $f = W/2 = 2.5~ {\rm GHz}$.}
\label{fig:edgeP}
\end{figure}

In our running example, we consider $K = 4$ stacked AMAF-RIS structures at height $h = 20$m 
on the ground, serving a sector on the ground x-y plan of the coordinate system S1 
with range between $r_{\rm min} = 10$m to $r_{\rm max} = 100$m 
azimuth $\phi$ from $-60\degree$ to $60\degree$. 
The 10m and 100m ground distances correspond to the downlook angles of $\alpha_{\rm max} = \text{acot}~ (r_{\rm min}/h) = 63.43\degree$ and $\alpha_{\rm min} = \text{acot}~ (r_{\rm max}/h) = 11.30\degree$, respectively, with respect to the origin of the system S2. 
Therefore, we choose the RIS mechanical downtilt angle $\alpha$ to be the arithmetic mean, i.e., $\alpha = 37.37 \degree$. 
This downtilt angle causes the RIS normal vector to intercept the ground at distance $r = 26.16$ m. Fig. \ref{fig:CenterSpot0} shows the ground footprint 
of the RIS PEM spot beam pointing at the RIS boresight (i.e., without any electronic steering) and at the center-bandwidth,
i.e., at baseband $f = 0$. 
We notice that thanks to the symmetric RIS excitation and the natural taper in the RIS amplitude profile said before, there are no sidelobes or energy 
spills beyond the  desired spot beam. 
In Fig. \ref{fig:edgeP}, we see that the ground illumination of the spot beam essentially remains the same at the edge frequencies (we show only the upper edge, to save space and for clarity of illustration), confirming that for this case the effect of frequency selectivity due to the propagation through the AMAF-RIS module is minimal (negligible beam ``squinting''). 

\begin{rem}
It is interesting to notice that in the real 3D setting with well-designed mechanical downtilt it is possible to ``focus'' 
the transmit energy on relatively restricted spotbeams both in angle and in range. Recently, there has been a lot of interest
in exploiting near-field propagation  conditions for {\em beam focusing}, exploiting the characteristics of the near-field propagation (e.g., see \cite{Near_Field_Beam_Focus1, Near_Field_Beam_Focus2}. The main idea is that while beamforming in the far-field
can focus the signal power only in the angle dimension, beam focusing in the near-field can focus the signal both in angle and in range, creating more 
spatial multiplexing opportunities for users with similar angle but separated in range. 
However, as it is known from \cite{liu2023near,bjornson2021primer}, the beamfocusing region is only about 10\% of the Fraunhofer distance. 
Hence, this beamfocusing effect is practically very limited. In addition,
this idea that beamforming in the far-field is not able to separate users with respect to the range is a {\em severe misconception} 
due to the fact that most such results are produced for 2D geometries and linear arrays. 
As it is apparent from our results, in 3D with planar arrays and mechanical/electronic downtilt it is perfectly possible to create localized spotbeams also
in the far-field. \hfill $\lozenge$
\end{rem}

\subsection{Link Budget}
\label{subsec:linkbudget}

In order to determine the transmit power, we consider the system requirement specifications of Table \ref{tab:SRS}. 
For the maximum cell range $r_{\rm max}=100$m, the maximum RIS-user slant distance (from the topmost RIS (sub-)array) is $\rho_{\rm max}=\sqrt{(h+3(h_{\rm sep}+N_z)\lambda/2)^2+r_{\rm max}^2}= \sqrt{20.015^2+100^2} = 101.98$m. Recalling $L_{\rm max} = (\lambda/4\pi \rho_{\rm max})^2$ is the maximum free-space pathloss at the maximum distance $\rho_{\rm max}$ (m), 
the 100 GHz carrier frequency and the maximum slant distance $\rho_{\rm max} = 101.98$m between a cell edge user and the base station, yield $L_{\rm max}~ ({\rm dB})=32.5 + 20~ \text{log}_{10} (100~{\rm GHz}) + 20~ \text{log}_{10} (102~{\rm m})=112.7~{\rm dB}$. For simplicity, we ignore the atmospheric attenuation which is in general less than a dB \cite{AAted}, \cite{ITUgasattn} depending upon the weather conditions. 
Recall that the thermal noise power spectral density is -174 dBm/Hz at 25\degree C, which is further raised by the receiver noise figure. 
Considering the $W=5$ GHz bandwidth (from Subsection \ref{subsec:T}) and the receiver (Rx) noise figure of 5 dB, the Rx noise power is $WN_0~({\rm dB}) = -174~ {\rm dBm/Hz}~ + 10~\text{log}_{10} (5~{\rm GHz})+5=-72~{\rm dBm}$. 
Let the required Rx signal to noise ratio (SNR) at the sector edge be 0 dB\footnote{Notice that an SNR of 0 dB corresponds to a 
channel capacity equal to 1 bit/s/Hz, which is approachable in practice using QPSK modulation with powerful binary LDPC coding of rate (slightly less than) 1/2.
Hence, such a system is quite realistic also from a practical viewpoint.}.
This means that the Rx signal power should be $-72~{\rm dBm}$. With the $L_{\rm max}=112.7~{\rm dB}$,
the effective isotropic radiated power (EIRP) should be $P_T = -72~{\rm dBm} + 112.7~{\rm dB} \approx 40.7~ {\rm dBm}$. 
It follows that the RF feed power from the AMAF, $P_{\rm amaf}=P_T/G(\phi,\theta)$\footnote{This implies that $\sigma_1 = 1$. In the numerical simulations, we get $\sigma_1=1.2$. A $\sigma_1>1$ is not practically possible (in hardware) because AMAF-RIS propagation is a passive mechanism. Therefore, we use $\sigma_1 = 1$ in this numerical study. Hardware-based determination of $\sigma_1$ is an extension work.}, 
where $G(\phi,\theta)$ is from \eqref{eq:bfpattern}. At the cell edge, $G(\phi=60\degree,\theta=26.06\degree)=18.7~{\rm dBi}$. We get $P_{\rm amaf}=40.7~{\rm dBm}-18.7~{\rm dBi}=22~{\rm dBm}=158.5~{\rm mW}$.

\begin{table}[h!]
\caption{Example system specifications.}
\label{tab:SRS}
\centering
\setlength{\tabcolsep}{2pt} 
\begin{tabular}{lclc}
\hline\hline 
Specification & Value & Specification & Value \\
\hline 
Carrier freq. (GHz)       & 100 & Receive noise pow. (dBm)     & -72   \\
Cell range (m)            & 10 to 100  & Receive SNR (dB)  & 0 \\
Azimuth span ($\phi$)     & +/-60$\degree$ & Receive signal power (dBm) & -72 \\
Bandwidth $W$ (GHz) & 5  &  Path Loss $L_{\rm max}$ (dB) & 112.7  \\
Thermal noise pow. (dBm)  & -77  & EIRP (dBm)    & 40.7  \\
Rx NF (dB)      & 5  & RIS size ($N_x\times N_z$)  & 16x16  \\
\hline 
\end{tabular}
\end{table}

\subsection{PEM Beam Pointing} 
\label{subsec:bpointing}

\begin{figure}[h!] 
\centerline{\includegraphics[width=8cm,height=4cm]{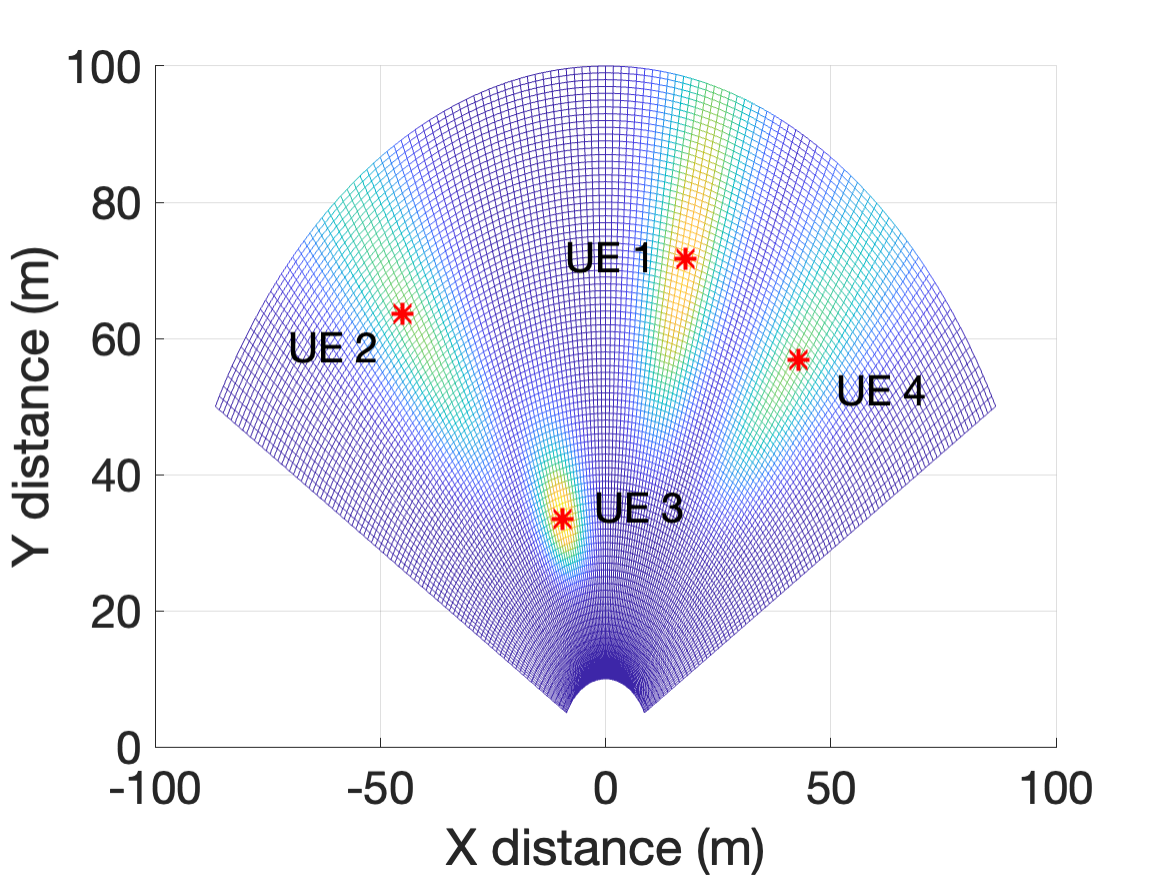}}
\caption{Ground footprints: an example set of 4 PEM spot beams, with perfect beam pointing onto the 4 UEs.}
\label{fig:MUspots}
\end{figure}

For the stacked multiuser  beamforming architecture, we choose the vertical separation between the RIS sub-arrays to be $h_{\rm sep} = 10$. 
With this choice,  the off-diagonal blocks in (\ref{eq:DTB2}) are very small. 
We consider a $K = 4$ downlink data streams serving 4 users randomly distributed with 
azimuth $\phi \in [-60\degree, 60\degree]$ and range $r \in [10\text{m}, 100\text{m}]$. 
We assume that the scheduler is ``smart'' and chooses the $K$ users with sufficient angular and distance separation 
on the coverage area.  In particular, in our results we imposed minimum azimuth angle separation between the users of $15\degree$, 
corresponding to the -20 dB beam contour. 
Fig. \ref{fig:MUspots} shows a snapshot (random realization) of 4 user positions and the corresponding 
ground beam footprints assuming perfect beam pointing, i.e., setting the electronic steering 
(RIS phase profile) such that, for each user $k$ at angle $(\phi_k, \theta_k)$, the corresponding RIS steering vector is  
$\wv_k = \av(\phi_k,\theta_k)$. The baseband matrix $|\Hm|^{\circ2}(0)$ (i.e., in power, and at the center frequency $f = 0$) 
from (\ref{eq:BB_MIMO}) corresponding to 
the example of Fig.~\ref{fig:MUspots} is given by 
\begin{equation}
\label{eq:PPH}
\centering
|\Hm|^{\circ2}(0) = \left [ \begin{array}{cccc} 
   ~26.4 & -21.7 & -55.4 & -6.8 \\
  -23.2 &  ~~25.1 & -7.9 & -30.9 \\
  -50.0 & ~-5.8  &  ~26.9 & -22.8 \\
   -8.1 & -30.5 & -25.0 &  ~~25.1 
\end{array} \right ] \text{dB}.
\end{equation}
Notice that the channel matrix is very strongly diagonally dominant. 
This has been observed in general, for all channel realizations provided that the minimum angular separation is imposed. 
With such a small multiuser interference, there is no need for baseband multiuser precoding (e.g., zero-forcing). 
Thus, the resulting system (pure beam steering) is also very simple to implement from the signal processing viewpoint. 
 
We also notice that, due to the directional patch element antenna factor 
$G_{\rm patch}(\phi,\theta)$, the useful signal term is attenuated for users at large azimuth angles with respect to the sector center $\phi = 0$
(in the example above, this can be noticed for users 2 and 4). In order to avoid this effect, the azimuth range should 
be restricted (e.g., to +/-45$\degree$ instead of +/-60$\degree$). Also, a smaller steering angle span enables a larger ISI-free bandwidth 
by (\ref{eq:tau}) and \eqref{eq:max_delay_spread}. For example, an angular span of +/-45$\degree$ allows 19$\%$ more ISI-free bandwidth than the angular span of +/-60$\degree$.

\begin{figure}[h!] 
\centerline{\includegraphics[width=8cm]{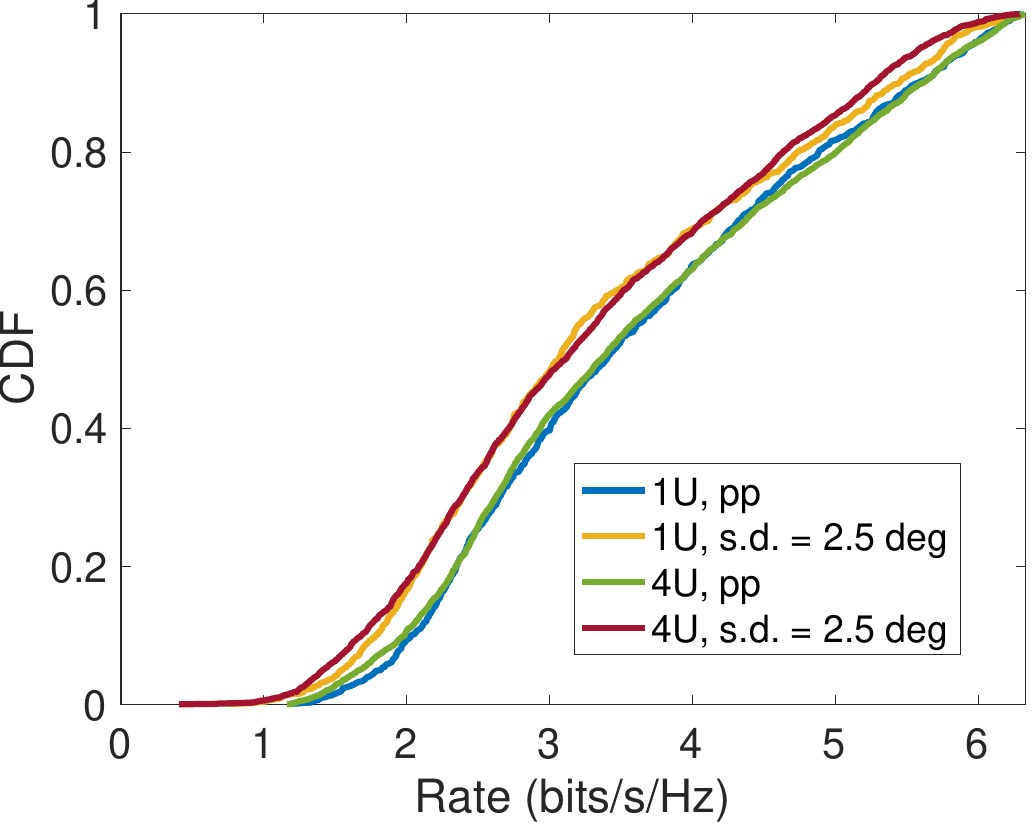}}
\caption{Rate CDFs for the beam pointing case, with perfect pointing (pp) and with Gaussian beam pointing errors.}
\label{fig:PP_rate_CDF}
\end{figure}

Fig. \ref{fig:PP_rate_CDF} shows the achievable rate CDF for the case of a single user (1U),  and 4 users (4U), with perfect beam pointing 
(pp), and Gaussian distributed independent beam pointing errors in both azimuth $\phi$ and elevation $\theta$ 
with standard deviation of $2.5 \degree$.  
First, we note that the degradation due to the beam pointing errors is more pronounced for lower rates than for the higher rates because angular errors have larger impact at farther ranges. Then, we see that there is no practical degradation (for both perfect pointing and pointing errors) for the 4 users case as compared to the single user case. This confirms the fact that with the designed multiuser BF scheme the residual multiuser interference is negligible in this range of practical SNR. 
This is a consequence of the proposed design, choosing a sufficiently large inter-sub-array vertical separation $h_{\rm sep} = 10$ 
half-wavelengths and the minimum azimuth angular separation of -20 dB beamwidth between the groups of simultaneously served users. 
We conclude that for such a system no digital baseband multiuser precoding is needed. 
Of course, these conclusions may be different for different geometries and system parameters. 

The user rate spread of the CDF in Fig. \ref{fig:PP_rate_CDF} is due to the combination of multiple effects: 
different slant distances between users and base station, lower RIS array gains with respect to the maximum possible $\Gamma$ 
for angles far from the direction $(\phi = 0, \theta = 0)$, i.e., the center of the spot beam of Fig.~\ref{fig:CenterSpot0}, and of course
the random beam pointing errors. 

\begin{figure}[h!]
\centerline{\includegraphics[width=8cm]{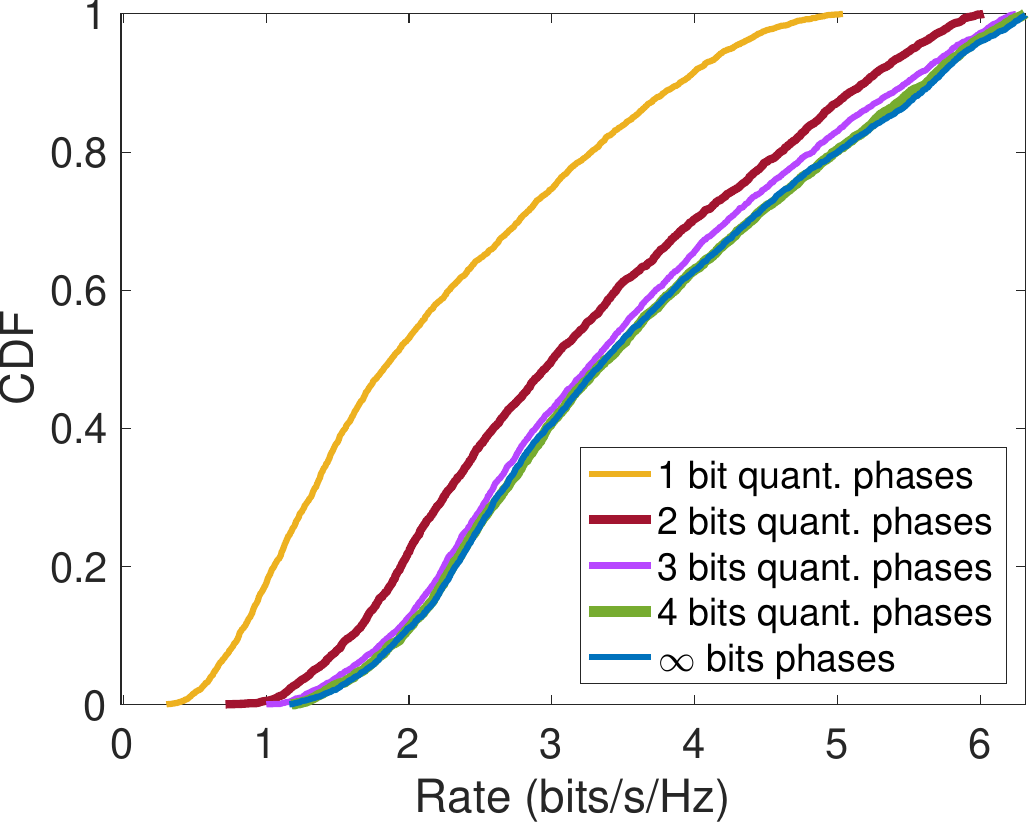}}
\caption{Rate CDFs for the beam pointing case, with perfect pointing but quantized phase-shifters at the RIS.}
\label{fig:Quant_MU}
\end{figure}

In Fig. \ref{fig:Quant_MU} we show the per-user rate CDF curves for the same scenario, considering 
1 bit, 2 bits, 3 bits, and 4 bits quantization of the phase-shifters at the RIS. 
We see that for  4 or more phase shift resolution bits the achievable rate CDF is essentially identical to
the case of unquantized phase-shifters. This show the robustness of the approach to practical hardware limitations such as
finte resolution in the phase shift control of the RIS. 

\subsection{PEM ``Naive'' Codebook Design} \label{subsec:cbook}

\begin{figure}[h!]
\centerline{\includegraphics[width=8cm,height=4cm]{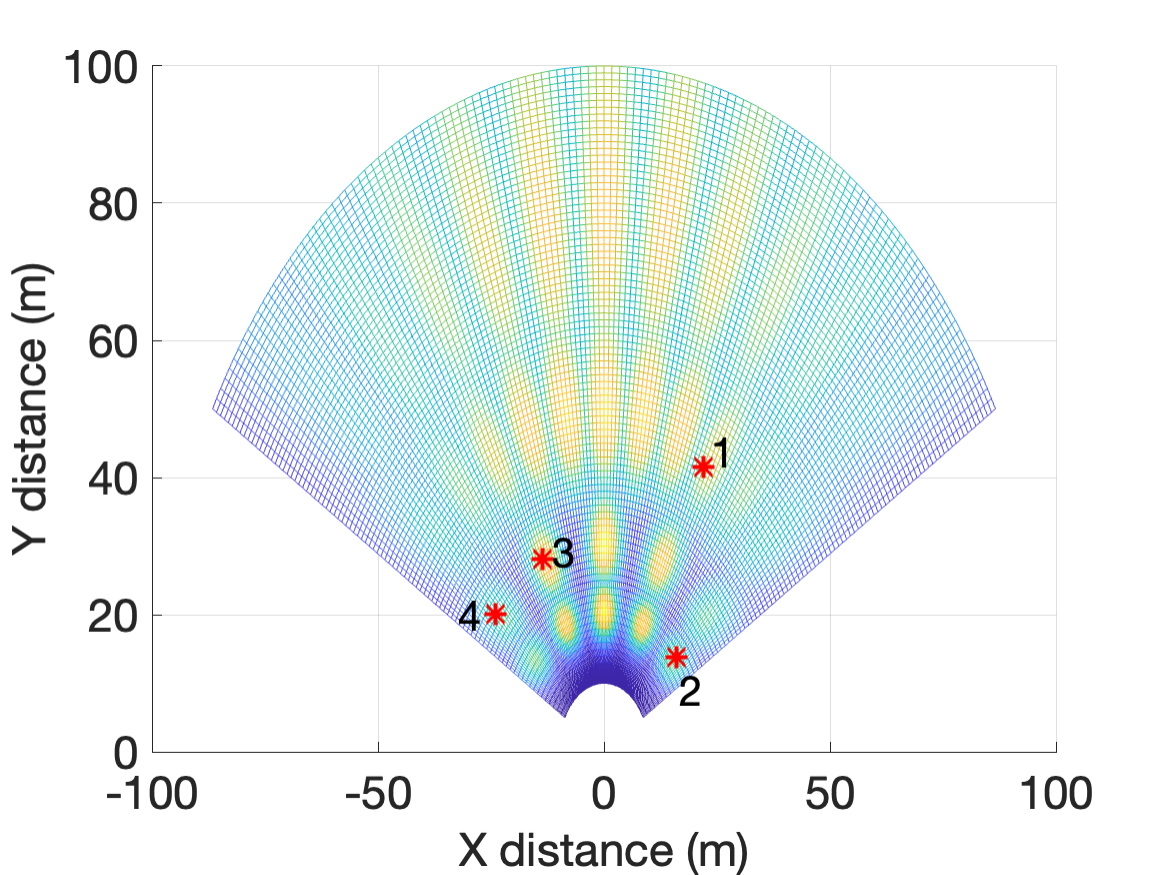}}
\caption{``Naive'' codebook ground footprint, with an example set of 4 UEs.}
\label{fig:CBook}
\end{figure}

\begin{figure}[h!]
\centerline{\includegraphics[width=8cm]{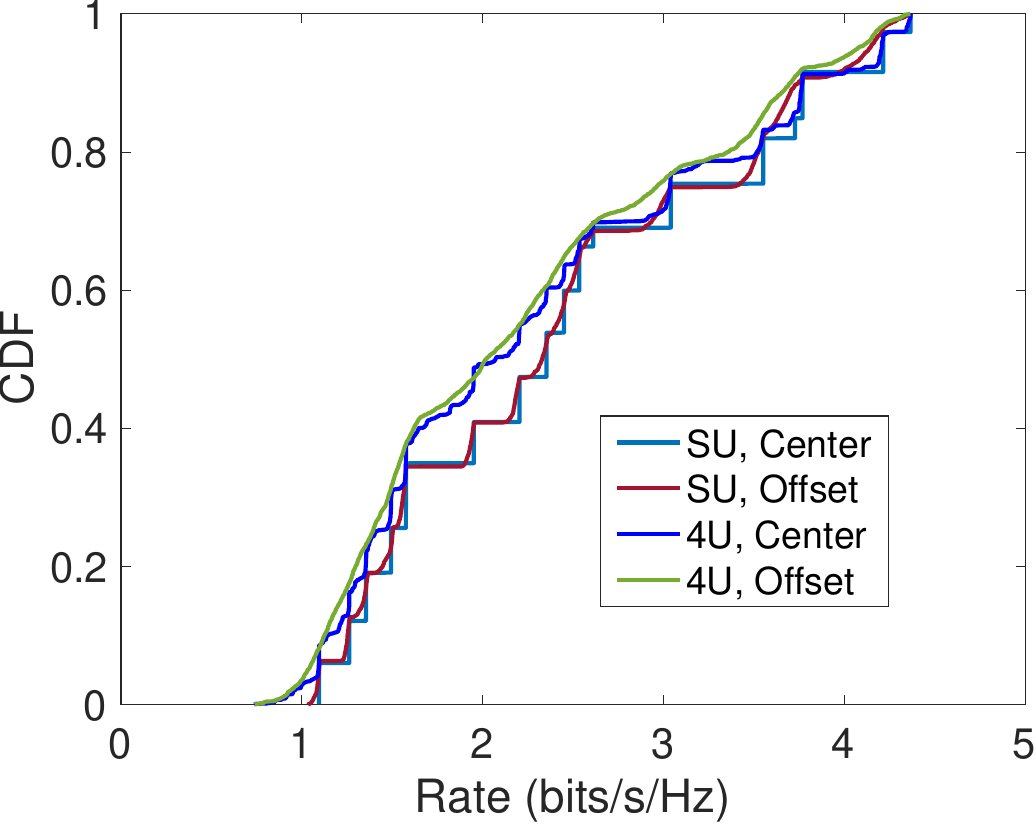}}
\caption{(``Naive'') Codebook-based approach: single user and multi-user rate CDFs with ideal and offset UE positions.}
\label{fig:CBookCDF}
\end{figure}

As an alternative to beam pointing (which requires user angle information, i.e., some form of positioning), a codebook based approach 
is widely considered practical wireless standards \cite{38.214}. 
In this case, a fixed set of beams is pre-designed in order to cover the desired coverage region. 
A standard beam acquisition (or beam ``sweeping'') scheme is used, such that the base station sends periodically probing signals (the so-called SSB signals in 
3GPP parlance \cite{38.213}) through its beams, such that the users can identify their most favorable beam and feed back the corresponding beam index to the base station. Then, the base station selects groups of users to be served on the same time slot using spatial multiplexing, 
making sure that the selected group of users correspond to sufficiently mutually separated beams. 
We have designed a beamforming codebook obtained from the 
PEM spot beam of Fig. \ref{fig:CenterSpot0}.
The footprints of a subset of beamforming codewords for 
four rings with ground distances of 20m, 30m, 50m, 80m  and angles 0, +/-25, +/-50 deg for the first two inner rings, then 0, +/-10, +/-20, +/- 30, +/- 40, +/-50 deg for the second two outer rings in shown in Fig. \ref{fig:CBook}. The whole codebook includes other subsets in order to offer an approximately uniform (i.e., without gaps) coverage.  They are not shown in the same figure for the sake of clarity.  Because this codebook is based only on ``naive'' electronic steering across the cell, we refer to it as a ``naive'' codebook design. $K = 4$ UEs are dropped at random into uniformly drawn 4 distinct beams with random offsets from respective beam centers.  
In Fig. \ref{fig:CBookCDF}, it is evident that the 4 user rate CDF is close to the single user rate CDF, but for a negligibly small degradation due to multiuser 
interference due to nearby BF codewords. Also in this case, further multiuser precoding in the baseband domain 
are not expected to yield significant gains. 

\section{Power Efficiency Analysis} 
\label{sec:PE}

\begin{figure*}
\centerline{\includegraphics[width=5in]{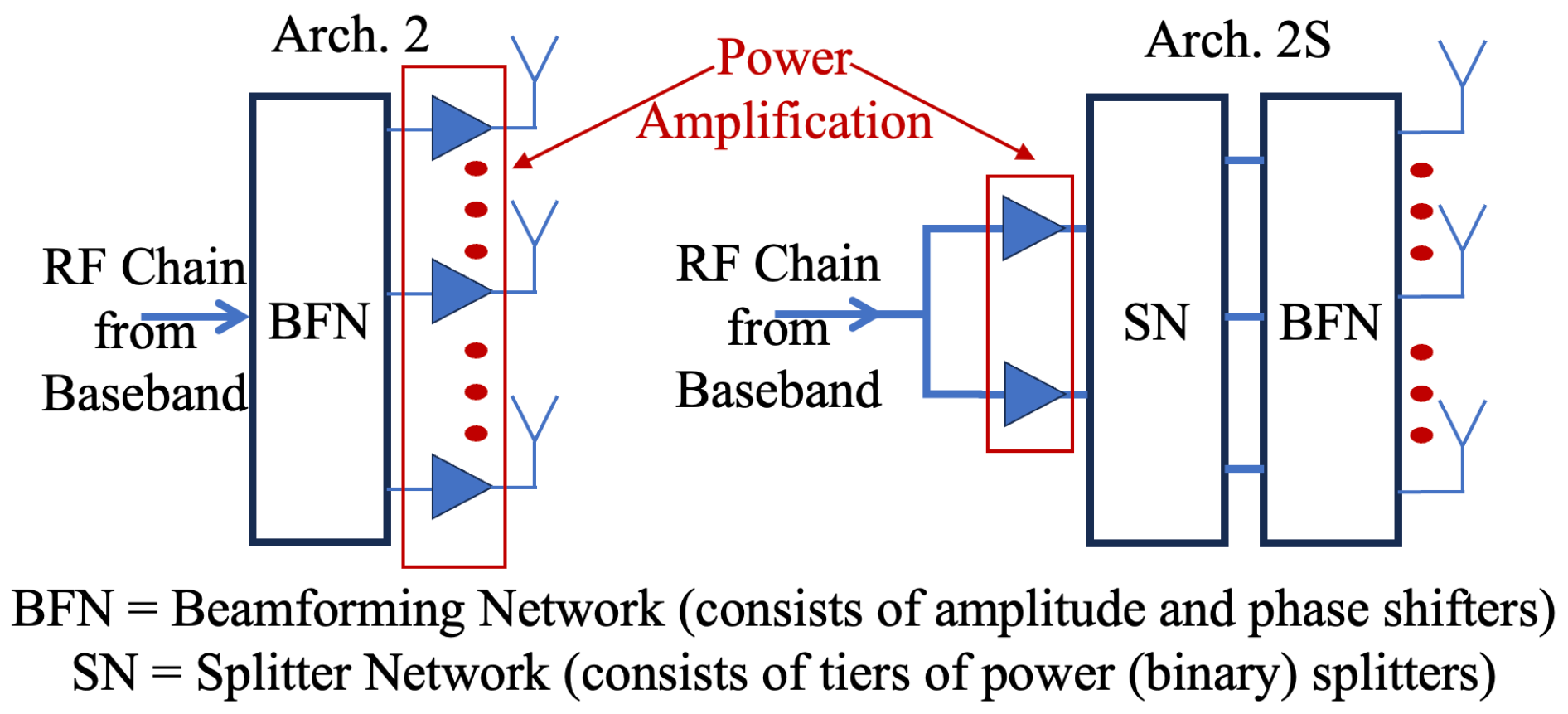}}
\caption{Block diagrams of classical architectures - Arch. 2 and Arch. 2S. BFN = Beamforming Network (consists of amplitude and phase shifters). SN = Splitter Network (consists of tiers of power (binary) splitters).}
\label{fig:Archs}
\centering
\end{figure*}

In \cite[Section V]{ICC2023}, a comparison of the power efficiency of principal eigenmode RIS beamforming versus conventional constrained-fed phased array technology was presented. The comparison considered the AMAF and the RIS configured as linear arrays. We now consider the full 3D beamforming, with the planar arrays. 

We consider the system requirement specifications in Table \ref{tab:SRS}. For the same beamforming performance of both our space-fed AMAF-RIS architecture (Arch. 1) and the classical constrained-fed active arrays (Arch. 2, where each element has its own dedicated power amplifier, see Fig. \ref{fig:Archs} on the top of the next page), we consider the same planar array dimensions with the same $|\uv_1|^{\circ2}$ taper profile as that of the PEM beamforming, we saw in Fig. \ref{fig:RIS_Ex}. Recall from Section \ref{sec:designs} that the AMAF and the RIS tapers in $|\vv_1|^{\circ2}$ and $|\uv_1|^{\circ2}$ are 11.3 dB and 58.9 dB, respectively. The 11.3 dB $|\vv_1|^{\circ2}$ taper is equivalent to the 58.9 dB $|\uv_1|^{\circ2}$ taper, in the sense both Arch. 1 with its active array (AMAF) taper $|\vv_1|^{\circ2}$ and Arch. 2 with its active array taper $|\uv_1|^{\circ2}$ have the same BF performance, thanks to the linear transformation of $\Tm(0)$ in Arch. 1.  

We also consider the case where a much smaller number of PAs feed an antenna array such that a group of antenna elements is constrained-fed by a power amplifier (PA) via a splitter network, e.g., Arch. 2S as shown in Fig. \ref{fig:Archs}, where the suffix `S' indicates the usage of a splitter network. The grouping of the antenna elements is manually optimized for the lowest possible DC power consumption. If a splitter network is used at the AMAF, we refer to it as Arch. 1S. We shall take-up these architectures for the power efficiency analysis in the following order: (i) Arch. 1, (ii) Arch. 1S, (iii) Arch. 2, and (iv) Arch. 2S. 

For a 1 to $\Nc$ power splitter stage with an insertion loss of $\Ic$, the input-to-per-output-port power ratio (for the stage)
\begin{equation}
\label{eq:IL}
\Lc  = \Ic\Nc.
\end{equation}
Notice that the splitter stage input-to-per-output-port power ratio, $\Lc\geq1$.
For a power amplifier (PA) feeding $N$ antenna elements with cumulative RF power $P_{\rm RF}$, having a certain power taper with the maximum coefficient $\Omega$, via an $N_s$ stage splitter network with a per-stage input-to-per-output-port power ratio of $\Lc$, the DC power consumption
\begin{equation}
\label{eq:PA.DC}
 \frac{P_{\rm DC}}{\Lc^{N_s}} = \frac{\Omega P_{\rm RF}}{\eta} \implies P_{\rm DC} = \frac{\Lc^{N_s} \Omega P_{\rm RF}}{\eta},
\end{equation}
where $\eta$ is the efficiency\footnote{PA efficiency actually includes a factor $\kappa$ which depends on several design details such as regime of linearity requested, PAPR of the modulation format, etc., which go beyond the scope of this paper. Therefore, for simplicity we assume $\kappa=1$. We neglect the small signal power consumption before the PA stage, which is practically reasonable with PA gain greater than 20 dB.} of the power amplifier.

We neglect the small signal power consumption before the PA stage which is practically reasonable as the small signal power consumption is relatively much smaller, and common to all the architectures. Further, we assume the same control power consumption for all the architectures, again for simplicity. We now do the numbers for the four architectures in the following paragraphs\footnote{Specific to our case study. This is because it is difficult to generalize the $|\vv_1|^{\circ2}$ and the $|\uv_1|^{\circ2}$ taper coefficient values for antenna elements groupings in Arch. 1S and Arch. 2S, respectively. In any case, the analysis methodology is applicable in general.} and present the consolidated DC power consumption values in Table \ref{tab:PC}. 

\textbf{Arch. 1}: From Subsection \ref{subsec:linkbudget}, the feeder RF power $P_{\rm amaf}=22~\text{dBm}$. Maximum element of $|\vv_1|^{\circ2}$, ${\rm max}|\vv_1|^{\circ2} =\Omega_{\rm amaf} = 0.154 = -8.12~\text{dB}$. Hence, for Arch. 1, the maximum RF requested is $P_{A1} = \Omega_{\rm amaf} \times P_{\rm amaf} = -8.12~\text{dB}+22~\text{dBm}=13.88~\text{dBm}=24.4~\text{mW}$. We assume that all the PAs in the (AMAF) array are developed in the same semiconductor technology, and are all biased with the same DC power dictated by the maximum requested RF power.  Considering Indium Phosphide (InP) PAs with efficiency $\eta=0.3$ \cite{ETH_PA_Survey, PAlimits},  the Arch. 1 DC power consumption, $P_{\rm 1} = N_a \Omega_{\rm amaf} P_{\rm amaf} / \eta  = 16 \times 24.4~\text{mW} / 0.3 = 1.3~ \text{W}$.

\textbf{Arch. 1S}: We group the 16 AMAF elements in two groups. Each group is fed by its own PA, via a 3 stage binary splitter network with 1 dB insertion loss per stage, (i.e., $N_s=3$ in \eqref{eq:PA.DC}, $\Ic= 1~ {\rm dB}$ and $\Lc=\Ic\Nc= 4~ {\rm dB}$), then the DC power consumption of the group 1 PA (the max RF, $\Omega_{\rm amaf}P_{\rm amaf}$=13.88 dBm), $P_{1S1} = 1.3~ \text{W}$, using \eqref{eq:PA.DC}.
Similarly, group 2 (max RF 8.27 dBm) requires DC power $P_{1S2} =  354.7~ \text{mW}$. Thus, the Arch. 1S DC power consumption, $P_{\rm 1S} = P_{1S1} + P_{1S2} = 1.7~ \text{W}$. 

\textbf{Arch. 2}: From the PEM $|\uv_1|^{\circ2}$ taper profile we saw in Fig. \ref{fig:RIS_Ex}, maximum element of $|\uv_1|^{\circ2}$, ${\rm max}|\uv_1|^{\circ2}=\Omega_{\rm ris}=-13.36~\text{dB}$, and ${\rm min}|\uv_1|^{\circ2} = -72.24~\text{dB}$. Therefore, Arch. 2 should be very ``power-hungry'' due to the 58.9 dB $|\uv_1|^{\circ2}$ power taper, and also the minimum RF power requested by an element is $P_{\rm amaf} ~{\rm min}(|\uv_1|^{\circ2})=21.3~\text{dBm}-72.2~\text{dB}\approx -50.2~\text{dBm}$, which is too low\footnote{Therefore, the 16x16 active array elements must be grouped as in Arch. 2S.}. The Arch. 2 DC power $P_{\rm 2} = N_p\Omega_{\rm ris}P_{\rm amaf}/\eta=6.2~ \text{W}$.

\textbf{Arch. 2S}: As an extreme case, an 8-stage binary splitter network, i.e., $N_s=8$, $\Nc=2$, $\Ic= 1~ {\rm dB}$, and $\Lc^{N_s}=32~{\rm dB}$, feeding all the $N_p=256$ elements with the maximum requested level of $P_{A2}=\Omega_{\rm ris}P_{\rm amaf}=8.64 $ dBm, then the DC power $P_{3a}=11.6/0.3=38.7$ W from \eqref{eq:PA.DC}.
Therefore, and also to obviate the need for the 58.9 dB dynamic range, and also to have a smaller number of splitter stages for lower power consumption, we must subgroup the antenna elements for smaller dynamic ranges. 

The antenna elements can be subgrouped in multiple ways. We present only the lowest possible DC power consumption case. Let 5 PAs feed 5 subgroups of the 256 elements with max RF power and the number of elements per subgroup as: (i) Subgroup 1: 8 elements with max RF $P_{A2}$, $N_s=3$, $\Nc=2$, $\Ic= 1~ {\rm dB}$, and $\Lc^{N_s}=12~{\rm dB}$, DC power $P_{\rm 2S1} =  386.3~ \text{mW}$. (ii) Subgroup 2: 8 elements with max RF 6.69 dBm, same splitter network, DC power consumption $P_{\rm 2S2} = 246.53~ \text{mW}$. (iii) Subgroup 3: 56 elements with max RF 3.91 dBm, 3 stage binary splitter network and a one-stage 1-7 splitter network (of 1 dB insertion loss), DC power consumption $P_{\rm 2S3} = 1.15~ \text{W}$. (iv) Subgroup 4: 56 elements  with max RF -5.68 dBm, 3 stage binary splitter network and a one-stage 1-7 splitter network (of 1 dB insertion loss), DC power consumption $P_{\rm 2S4} =  125.9~ \text{mW}$. (v) Subgroup 5: the remaining 128 elements with max RF -16.24 dBm, 7 stage binary splitter network, DC power consumption $P_{\rm 2S5} = 50~ \text{mW}$. Thus, the Arch. 1S DC power consumption, $P_{\rm 2S} = P_{\rm 2S1} + P_{\rm 2S2} +P_{\rm 2S3} +P_{\rm 2S4} +P_{\rm 2S5} = 1.96~\text{W}$.  

\begin{table}[h!]
\caption{A power comparison of different architectures.}
\label{tab:PC}
    \centering
    \setlength{\tabcolsep}{2pt}
    \begin{tabular}{clc}
    \hline\hline 
    Sl. No. & Architecture & Power consumption (W) \\
    \hline 
      1   &  Arch. 1 ($P_{\rm 1}$)& 1.3 \\
      2   & Arch. 1S ($P_{\rm 1S}$) & 1.7\\
      3   & Arch. 2 ($P_{\rm 2}$)& 6.2\\
      4   & Arch. 2S ($P_{\rm 2S}$) & 2.0\\
      \hline 
    \end{tabular}
\end{table}

From the consolidated values in Table \ref{tab:PC}, we see that Arch. 1 is the most power efficient. Finally, it should be noted that the above-mentioned power efficiency of the AMAF-RIS over-the-air eigenbeamforming is achieved in addition to the greatly simplified hardware requirements compared to conventional guided-wave feeding of a large phased-array or hybrid beamforming architecture, which requires a very complex feeding network (256 elements with 59 dB power taper!). Therefore, the proposed AMAF-RIS architecture also excels from the thermal management and reliability viewpoints.  Furthermore, multiple power amplifiers of the AMAF can be used for RF power build-up, which is otherwise not possible, for example, with a single PA driven feed horn. This not only allows for much greater ranges due to the much greater combined RF power from the multiple PAs to the AMAF, but also provides the feature of graceful degradation in the event of one or more power amplifier failures. Considering all of the above advantages, our novel eigenbeamforming AMAF-RIS designs enable a high-potential and agile communication system for both terrestrial networks and non-terrestrial networks (NTNs) envisioned for 6G and beyond, e.g., it will drastically reduce not only the test and space qualification efforts, but also the size, weight, power, and cost (SWaP-C) of satellites with very high EIRP and on-the-fly beamforming and steering capabilities.

\section{Conclusions}  
\label{sec:CONC}

We have proposed a novel multiuser, multibeam AMAF-RIS principal-eigenmode based over-the-air beamforming design, suited to very high frequency bands, 
LOS propagation, and achieving remarkable performance with high energy efficiency and low hardware complexity. 
The key idea is to use a RIS in the near-field of a small active array (the AMAF) in order to perform electronically steerable beamforming 
with high gain and directivity, while avoiding the complexity and power inefficiency of large active arrays. The far-field planar wave propagation delay difference across the AMAF-RIS structure, and the resulting frequency-selectivity, is tackled using only a few OFDM subcarriers (orders of magnitude fewer than the number of subcarriers required for much larger RIS apertures necessary in far-fields).
This design can provide high data rates at ranges up to 100s of meters for mobile access. 
The principal eigenmode precoding at the AMAF with a properly chosen AMAF-RIS distance such that the AMAF is in the near field of the RIS achieve
remarkable energy efficiency and very desirable taper of the amplitude profile at the RIS elements, with a very limited taper (dynamic range) at the AMAF active 
antennas. In turn, the RIS taper yields beamforming with very low side lobes. 
We provided a detailed analysis of the power efficiency of our architecture compared with alternatives based on more standard active array design. 
We examined a multiuser MIMO communication case study and verified that moderate beam pointing errors and suitable 
quantization of the RIS phase shifters do not incur dramatic performance losses.  
Future research directions include: (i) hardware-based validations of the idea, and (ii) joint optimization of the RIS phase-shifters and the AMAF 
precoder as an alternative to the simple PEM design. 

\appendix

In \cite[Tables I and II, Fig. 6]{ICC2023}, the RIS gain $\Gamma$ remains essentially constant 
with increasing RIS size. This is due to the fact that in \cite{ICC2023} we considered a 2D geometry with RIS and AMAF implemented by 
ULAs. To understand the scaling of the RIS gain $\Gamma$ with the RIS aperture in the 3D geometry of this paper, 
let us consider a ratio $F/D=1$. 
For a squared $N \times N$ SRA RIS, the largest RIS dimension is $\text{D} = \sqrt{2}N \lambda/2$. 
Letting $\lambda=1$ for simplicity, this results in $\text{D}= N/\sqrt{2}$. The AMAF-RIS path loss 
\begin{equation}
\label{eq:a1}
 L_R = (4 \pi F \lambda/2)^2 / \lambda^2 = 4 \pi^2 F^2,
\end{equation}
The RIS aperture is $A \propto D^2=N^2/2$. 
As $N$ increases, $F$ also increases linearly with $N$ considering the constant ratio $F/D =1$. 
Thus, we have
\begin{equation}
\label{eq:a2}
 L_R \propto N^2.
\end{equation}
On the other hand, the aperture gain is proportional to the physical aperture, i.e., 
\begin{equation}
\label{eq:a3}
G \propto A = N^2/2.
\end{equation}
Therefore, the RIS capture of inward radiated power from the AMAF is also proportional to $N^2$, which means that for increasing 
$N$ and constant ratio $F/D$, the increased AMAF-RIS loss $L_R$ is compensated by the correspondingly increased inward RIS aperture gain.
In other words, a larger RIS aperture is proportionally farther apart from the AMAF so that it collects the same electromagnetic energy. 

However, a larger RIS can focus the captured electromagnetic energy into a narrower beam in the far-field, where (\ref{eq:a3}) applies again 
(this time for the ``outward'' reflected radiation), Therefore, the overall AMAF-RIS gain is
\begin{equation}
\label{eq:a4}
\Gamma \propto \text{A} = N^2/2.
\end{equation}

\begin{table}[ht]
\centering
\caption{RIS gain $\Gamma$ vs. RIS size $N$.}
\label{tab:RIS_Gain}
\begin{tabular}{ccc}
\hline\hline
$N$  & f & $\Gamma$ (dBi) \\
\hline 
16  & 8  & 25.6 \\
32  & 16 & 31.1 \\
64  & 30 & 36.4 \\
128 & 80 & 44.7 \\ \hline 
\end{tabular}
\end{table}

As seen in Table \ref{tab:RIS_Gain} (for $N \times N$ SRA RIS aperture), the numerical computation of the AMAF-RIS gain confirms the linear growth 
with the planar array size, i.e., with $N^2$. There is a small deviation from the strict linearity, which depends on the specific $F/D$ ratio to achieve a specific taper 
profile. The linear increase of the gain with the array size is analogous to the classical parabolic reflector with feeder located at the focus, 
for which it is well known that the gain grows linearly with the reflector aperture. 

Finally, for the 2D geometry and the linear RIS considered in \cite{ICC2023}, the RIS aperture gain is
\begin{equation}
\label{eq:a5}
\text{G}_{\rm Linear} \propto N_p.
\end{equation}
The combination of energy capture and energy focusing of the linear
RIS gives a combined gain that grows with $N_p^2$.
With constant $F/D$, and $D \propto N_p$, the AMAF-RIS loss also grows with $N_p^2$. Thus, the $N_p^2$ increase in the RIS gain compensates for the $N_p^2$ increase in the AMAF-RIS loss. 
This explains why, in the 2D geometry of \cite{ICC2023}, we found that $\Gamma_{\rm Linear}$ remains essentially constant
as the RIS size $N_p$ increases. 

\bibliographystyle{IEEEtran}
\bibliography{EMP-bibliography}

\end{document}

%% file: macros.tex
\newfont{\bbb}{msbm10 scaled 700}

\newfont{\bb}{msbm10 scaled 1100}
\newcommand{\CC}{\mbox{\bb C}}

\newcommand{\RR}{\mbox{\bb R}}

\newcommand{\av}{{\bf a}}
\newcommand{\bv}{{\bf b}}

\newcommand{\iv}{{\bf i}}
\newcommand{\jv}{{\bf j}}
\newcommand{\kv}{{\bf k}}

\newcommand{\nv}{{\bf n}}

\newcommand{\pv}{{\bf p}}
\newcommand{\qv}{{\bf q}}

\newcommand{\uv}{{\bf u}}
\newcommand{\wv}{{\bf w}}
\newcommand{\vv}{{\bf v}}
\newcommand{\xv}{{\bf x}}

\newcommand{\onev}{{\bf 1}}

\newcommand{\Am}{{\bf A}}
\newcommand{\Bm}{{\bf B}}

\newcommand{\Hm}{{\bf H}}

\newcommand{\Rm}{{\bf R}}

\newcommand{\Tm}{{\bf T}}
\newcommand{\Um}{{\bf U}}
\newcommand{\Wm}{{\bf W}}
\newcommand{\Vm}{{\bf V}}
\newcommand{\Xm}{{\bf X}}

\newcommand{\Ic}{{\cal I}}

\newcommand{\Lc}{{\cal L}}

\newcommand{\Nc}{{\cal N}}

\newcommand{\Sigmam}{\hbox{\boldmath$\Sigma$}}

\newcommand{\Thetam}{\hbox{\boldmath$\Theta$}}

\newcommand{\diag}{{\hbox{diag}}}

\newcommand{\herm}{{\sf H}}

\newcommand{\transp}{{\sf T}}

\usepackage{stmaryrd} 

\usepackage{hyperref}
\hypersetup{
    bookmarks=true,         
    unicode=false,          
    pdftoolbar=true,        
    pdfmenubar=true,        
    pdffitwindow=false,     
    pdfstartview={FitH},    
    pdfnewwindow=true,      
    colorlinks=true,       
    linkcolor=red,          
    citecolor=green,        
    filecolor=blue,      
    urlcolor=blue           
}